\documentclass[aps,twocolumn,prb,preprintnumbers,superscriptaddress,amsmath,floatfix]{revtex4}
\usepackage{listings}
\usepackage{verbatim}

\usepackage{graphicx}
\usepackage{dcolumn}
\usepackage{bm}
\usepackage{soul}
\usepackage{color}
\usepackage{xcolor}
\usepackage{amsmath}
\usepackage{amssymb}
\usepackage{amsmath}
\usepackage{float}
\usepackage{natmove}
\usepackage{multirow}
\usepackage{setspace}
\usepackage{array}
\usepackage{booktabs}
\usepackage{rotating}
\usepackage{ulem}
\usepackage[colorlinks=true,linkcolor=blue,citecolor=blue,urlcolor=blue]{hyperref}
\usepackage{}
\usepackage{physics}
\usepackage{mathrsfs}
\usepackage{cprotect}
\usepackage{mathtools}
\usepackage[capitalise]{cleveref}
\usepackage[version=3]{mhchem} 
\definecolor{Grey}{rgb}{0.50,0.50,0.50}
\definecolor{Blu}{rgb}{0.00,0.00,1.00}
\definecolor{Red}{rgb}{1.00,0.00,0.00}
\definecolor{Green}{rgb}{0.20,0.50,0.20}
\definecolor{Magenta}{rgb}{0.60,0.00,0.60}
\definecolor{BluBondi}{rgb}{0.00,0.58,0.71}
\definecolor{Orange}{rgb}{0.95,0.46,0.17}
\definecolor{Red}{rgb}{1.00,0.00,0.00}

\newcommand{\editor}[2]{%
  \expandafter\newcommand\csname #1note\endcsname[1]{%
    \textcolor{#2}{{\it (\textbf{#1:} ##1)}}}%
  \expandafter\newcommand\csname #1\endcsname[1]{%
    \textcolor{#2}{##1}}%
  \expandafter\newcommand\csname #1cancel\endcsname[1]{%
    \textcolor{#2}{\sout{##1}}}%
  \expandafter\newcommand\csname #1change\endcsname[2]{%
    \textcolor{#2}{\sout{##1} ##2}}%
  \newenvironment{#1text}{\color{#2}}{\color{black}}
}

\editor{MB}{Green}
\editor{EM}{Red}
\editor{DP}{BluBondi}
\editor{AF}{cyan}
\editor{DV}{orange}

\newcommand{\suppinfo}{Supplemental Material~\cite{supp-info}}
\newcommand{\drop}[1]{}

\renewcommand{\emph}{\textit}

\begin{document}

\title{Towards high-throughput many-body perturbation theory: efficient algorithms and automated workflows}

\author{Miki Bonacci}
\email[corresponding author:]{miki.bonacci@nano.cnr.it}
\address{FIM Department, University of Modena and Reggio Emilia, Via Campi 213/a, Modena, Italy}
\address{S3 Center, Istituto Nanoscienze, CNR, Via Campi 213/a, Modena, Italy}
\author{Junfeng Qiao}
\address{Theory and Simulation of Materials (THEOS) and National Centre for Computational Design and Discovery of Novel Materials (MARVEL), \'Ecole Polytechnique F\'ed\'erale de Lausanne, CH-1015 Lausanne, Switzerland}
\author{Nicola Spallanzani}
\address{S3 Center, Istituto Nanoscienze, CNR, Via Campi 213/a, Modena, Italy}
\author{Antimo Marrazzo}
\address{Dipartimento di Fisica, Universit\`a di Trieste, I-34151 Trieste, Italy}
\author{Giovanni Pizzi}
\address{Theory and Simulation of Materials (THEOS) and National Centre for Computational Design and Discovery of Novel Materials (MARVEL), \'Ecole Polytechnique F\'ed\'erale de Lausanne, CH-1015 Lausanne, Switzerland}
\address{Laboratory for Materials Simulations (LMS), Paul Scherrer Institut (PSI), CH-5232 Villigen PSI, Switzerland}
\author{Elisa Molinari}
\address{FIM Department, University of Modena and Reggio Emilia, Via Campi 213/a, Modena, Italy}
\address{S3 Center, Istituto Nanoscienze, CNR, Via Campi 213/a, Modena, Italy}
\author{Daniele Varsano}
\address{S3 Center, Istituto Nanoscienze, CNR, Via Campi 213/a, Modena, Italy}
\author{Andrea Ferretti}
\address{S3 Center, Istituto Nanoscienze, CNR, Via Campi 213/a, Modena, Italy}
\author{Deborah Prezzi}
\address{S3 Center, Istituto Nanoscienze, CNR, Via Campi 213/a, Modena, Italy}

\date{\today}
\begin{abstract}
The automation of ab initio simulations is essential in view of performing high-throughput (HT) computational screenings oriented to the discovery of novel materials with desired physical properties. 
In this work, we propose algorithms and implementations that are relevant to extend this approach beyond density functional theory (DFT), in order to automate many-body perturbation theory (MBPT) calculations. 
Notably, a novel algorithm pursuing the goal of an efficient and robust convergence procedure for GW and BSE simulations is provided, together with its implementation in a fully automated framework. This is accompanied by an automatic GW band interpolation scheme based on maximally-localized Wannier functions, aiming at a reduction of the computational burden of quasiparticle band structures while preserving high accuracy. 
The proposed developments are validated on a set of representative semiconductor and metallic systems.
\end{abstract}
\maketitle

\section{Introduction}

Computational HT screening is nowadays a consolidated approach to materials discovery~\cite{curtarolo_high-throughput_2013,vecchio_high-throughput_2021,luo_highthroughput_2021}, as a complementary and accelerated tool with respect to experimental efforts. In the last decade, seminal works in this field have addressed, among many other topics, the discovery of novel 2D materials~\cite{ashton_topology-scaling_2017,cheon_data_2017,Mounet_2018,choud_2d,MC2DB,haastrup_computational_2018,marrazzo_relative_2019}, the identification of optimal new lithium-ion battery anodes~\cite{kirklin_high-throughput_2013,zhang_computational_2019}, thermoelectric~\cite{chen_understanding_2016,bhattacharya_high-throughput_2015}, photocatalysts~\cite{castelli_computational_2012} and photovoltaic light harvesting~\cite{yu_identification_2012,yan_solar_2017,kuhar_high-throughput_2018} materials.
The success of these studies relies on the development of different software and implementations~\cite{lejaeghere_reproducibility_2016} that were able to encode complex domain-specific knowledge into automated and robust workflows -- enforcing rigorous computational protocols~\cite{lejaeghere_reproducibility_2016,SSSP} and managing all the steps concerning a simulation -- which thus require the least possible human intervention~\cite{maffioletti_gc3pie:_2012,curtarolo_aflow:_2012,jain_fireworks:_2015,PIZZI2016218,hjorth_larsen_atomic_2017,mathew_atomate:_2017,mortensen_myqueue:_2020,huber_aiida_2020,uhrin_workflows_2021}. 

Concerning the electronic structure field, most of these works and implementations are based on DFT, which allows one to compute total energies, optimized geometries, and other ground state properties of materials with predictive accuracy.
However, different approaches are required for the accurate prediction of excited-state properties of materials, such as quasiparticle (QP) band structures and absorption spectra, which are typically crucial for the description of active processes in modern optoelectronic technologies, like photovoltaics, photocatalysis, light-emitting diodes (LEDs), photodetectors and solar cells~\cite{graph_opto,MXenes_opto_review,pathways_photovoltaics,zhu_fotoc,photodetectors_1D_CdSe,xia_ultrafast_2009,lee_highdetectivity_2019}.
In this context, MBPT and Green's function methods represent the state-of-the-art tools, where charged (electronic quasi-particle levels) and neutral excitations (optical properties, electron energy loss spectra) can be obtained by means of the GW approximation and the Bethe-Salpeter equation (BSE), respectively~\cite{martin_reining_ceperley_2016}.

To date, a limited number of attempts have been made toward automation~\cite{van_setten_automation_2017} and HT screening~\cite{haastrup_computational_2018,rasmussen_towards_2021} based on these MBPT approaches, mainly because of their conceptual and computational complexity. Indeed, depending on the specific physical problem, different levels of theory might be adopted~\cite{huser_quasiparticle_2013}, further branching off depending on the chosen approximations and implementations~\cite{rangel_repr_2020,stankovski_g_2011}. Furthermore, even for the simplest approximations, these calculations require the control over a much larger parameter space with respect to DFT, with parameters that are often interdependent and might change depending on the specific implementations adopted, but the choice of which is always crucial to obtain reliable results. From a computational point of view, this type of simulations are constituted by a chain of distinct steps, often relying on the usage of different software tools, each of them with their own specificity, e.g., in terms of memory and parallelization requirements. Memory requirements are much heavier than in standard DFT simulations even for moderate system size, and often require massive usage of parallel computing resources. Calculations often fail due to memory overflow and have to be restarted with careful choice of parameters. All of these problems make the application of MBPT-based approaches a complex and difficult task per se, and its automation still an open challenge. 

Building on pioneering works in the field~\cite{van_setten_automation_2017,rasmussen_towards_2021}, we here focus on the development of an improved algorithm aiming at an efficient and computationally cost-effective management of the choice of converged parameters for accurate GW-BSE calculations. The algorithm is implemented in the AiiDA framework~\cite{huber_aiida_2020,uhrin_workflows_2021}, a platform that is routinely used for HT studies \cite{Mounet_2018,mercado_silico_2018,prandini_precision_2018,Vitale2020} and that incorporates the ADES model for Automation, Data, Environment and Sharing~\cite{PIZZI2016218}. As detailed below, this implementation allows us to encode an efficient error handling, memory and parallelization management, and logic computational flows within automated python workflows. Moreover, it guarantees a seamless interoperability of different software codes that tackle the different steps usually involved in MBPT simulations, i.e., the preliminary DFT part (here using \textsc{Quantum ESPRESSO}~\cite{2009qe,2017qe}), the GW-BSE calculations (\textsc{Yambo}~\cite{yambo1,yambo2}), and any required post-processing. In particular, for the latter, we here introduce a scheme based on maximally localized Wannier functions~\cite{marzari_maximally_2012} for the automatic interpolation of GW band structures, which interfaces the \textsc{Wannier90}\cite{Pizzi2020} and \textsc{Yambo} projects. 
All of these developments point to a drastic reduction of both human and computational efforts, key issues for enabling HT studies. In addition, by incorporating the different domain-specific scientific and computational competences into robust and reliable workflows, we aim at making accurate GW-BSE calculations available for the materials science community at large, including non-experts in the field (e.g., via graphical user interfaces~\cite{aiidalab}), similarly to what has already happened with DFT.

The manuscript is organised as follows. In the Results and Discussion Section, we first introduce a model for the convergence surface in the N-dimensional space of the GW (BSE) variables, and present our improved algorithm for efficiently retrieving converged values for the main (interdependent) parameters. The implementation of this algorithm within the \texttt{aiida-yambo }plugin is then described, followed by the presentation of the novel \texttt{aiida-yambo-wannier90} plugin that encodes the GW band interpolation based on Wannierization. Both these implementations are validated for selected prototypical systems. Additional details on simulations are provided in the Methods section. In the remainder of this section, we instead introduce the main concepts and quantities related to the GW and BSE schemes, which will be useful to properly understand the subsequent Sections.  

\subsection*{GW approximation}

Accurate electronic band structures of materials can be computed within the MBPT framework by correcting the Kohn-Sham (KS) DFT eigenvalues with a self-energy term $\Sigma$ by means of the $GW$ approximation, i.e. $\Sigma$ is approximated with the first term of the perturbation series expansion in terms of the screened Coulomb interaction $W$~\cite{Hedin_1965}. We hereafter consider the simplest and most widespread implementation of GW, i.e. the so-called $G_0W_0$ approach, where $G_0$ is the KS independent-particle one-body Green's function and $W$ is computed within the random-phase approximation (RPA)~\cite{onida_electronic_2002}. Nevertheless, the convergence algorithm presented here can be used also for more sophisticated flavours of the theory, e.g. the self-consistent GW.
Under these assumptions and by considering a plane-wave expansion, the self-energy term $\Sigma_{n\textbf{k}}$ for a band $n$ at a given {\textbf{k}-point} -- written as the sum of the Fock exchange ($\Sigma^x$) and the frequency-dependent correlation ($\Sigma^c$) terms -- is given by:
\begin{equation}
\label{Sigma_x_yambo}
    \Sigma^x _{n\textbf{k}} = - \sum_m^{\text{occ}} \int \frac{d\textbf{q}}{(2\pi)^3} \sum_{\textbf{G}}^{G^x_{\text{cut}}} v_{\bf{G}}(\textbf{q}) |\rho_{nm}(\textbf{k},\textbf{q},\textbf{G})|^2f_{m,\textbf{k}-\textbf{q}}
\end{equation}
and
\begin{equation}
\label{Sigma_c_yambo}
\begin{split}
    \Sigma^c _{n\textbf{k}}(\omega) = &-i\sum_m^{N_b} \int  
    \frac{d\textbf{q}}{(2\pi)^3} \sum_{\textbf{G}\textbf{G}'}^{G_{\text{cut}}} \rho_{nm}(\textbf{k},\textbf{q},\textbf{G})\rho^*_{nm}(\textbf{k},\textbf{q},\textbf{G}') \times \\
    & \int d\omega^{\prime} W_{\bf{GG'}}(\textbf{q},\omega ')  \times \\ 
    &\Bigg[ \frac{f_{m,\textbf{k}-\textbf{q}} }{\omega - \omega' - \epsilon_{m,\textbf{k}-\textbf{q}}-i\eta} + \frac{1-f_{m,\textbf{k}-\textbf{q}} }{\omega - \omega' - \epsilon_{m,\textbf{k}-\textbf{q}}+i\eta} \Bigg],
\end{split}
\end{equation}
where $v_{\bf{G}}(\textbf{q})$ and $W_{\bf{GG'}}(\textbf{q},\omega ')$ are the bare and screened Coulomb interaction, respectively. The generalized dipole matrix elements $\rho_{nm}(\textbf{k},\textbf{q},\textbf{G})$ are defined as:
\begin{equation}
    \rho_{nm}(\textbf{k},\textbf{q},\textbf{G}) = \bra{n\textbf{k}}e^{i(\textbf{q}+\textbf{G})\cdot \textbf{r}}\ket{m\textbf{k}-\textbf{q}},
    \label{yambo_dip}
\end{equation}
where $\epsilon^{\text{KS}}_{n\textbf{k}}$ and $|n\mathbf{k}\rangle$ are the corresponding KS eigenvalues and eigenvectors. 
Here, $W$ can be written in terms of the reducible polarizability $\chi$ ($W = v + v \chi v$), which is in turn computed by solving a Dyson equation for the RPA irreducible polarizability $\chi^0$:
%
\begin{equation}
    \chi_{\bf{G}\bf{G}^\prime}
  (\textbf{q},\omega) = \sum^{G_{\text{cut}}}_{\bf{G}^{\prime \prime}} \left[I - v(\bf{q})\chi^0 (\bf{q},\omega)\right]_{\bf{G}\bf{G}^{\prime \prime}}^{-1}\chi^0_{\bf{G}^{\prime \prime}\bf{G}^{\prime }} (\bf{q},\omega),
\label{chi_dyson}
\end{equation}
where
\begin{widetext}
\begin{equation}
\label{chi_not_tensor}
    \chi^0_{\textbf{G}\textbf{G}'} (\textbf{q}, \omega) = 
    2 \sum_{nm}^{N_b}\int_{BZ} \frac{d\textbf{k}}{(2\pi)^3}  \text{ } \rho^*_{mn}(\textbf{k},\textbf{q},\textbf{G}) \rho_{mn}(\textbf{k},\textbf{q},\textbf{G}')  
    \left[ \frac{f_{n\textbf{k}-\textbf{q}}(1-f_{m\textbf{k}})}{\omega + \epsilon_{n\textbf{k}-\textbf{q}} - \epsilon_{m\textbf{k}}} - \frac{f_{n\textbf{k}-\textbf{q}}(1-f_{m\textbf{k}})}{\omega+ \epsilon_{m\textbf{k}} - \epsilon_{n\textbf{k}-\textbf{q}}} \right].
\end{equation}
\end{widetext}
In practice, the above quantities are computed by introducing specific approximations that crucially impact both the accuracy and the computational and memory costs of the calculations. In particular, Eq.~\eqref{chi_dyson} as well as Eq.~\eqref{Sigma_c_yambo} involve a discrete, finite number of $\textbf{G}$ vectors, set by a cutoff parameter $G_{\text{cut}}$, which defines the size of the response matrix. 
Moreover, with the parameter $N_b$, we introduce a truncation over the KS states summation for both response $\chi^0_{\textbf{G}\textbf{G}'}$ (Eq.~\eqref{chi_not_tensor}) and self-energy $\Sigma^c$ (Eq.~\eqref{Sigma_c_yambo}), which should in principle include all occupied and an infinite number of empty states.
Finally, all the integrals in reciprocal space are computed on a discrete \textbf{k}-points (\textbf{q}-points) grid whose size, $N_{\textbf{k}}$, defines the accuracy of the sampling of the Brillouin zone (BZ). 
All of these three parameters, $G_{\text{cut}}$, $N_b$, and $N_{\textbf{k}}$, need to be increased till the desired convergence is reached. 

One of the major obstacles to automate the convergence procedure lies in the interdependence of the first two parameters, $G_{\text{cut}}$ and $N_b$, such that their convergence has to be performed \textit{jointly}, as thoroughly discussed elsewhere~\cite{stankovski_g_2011,gao_speeding_2016,van_setten_automation_2017}.  
Indeed, \cref{chi_dyson,chi_not_tensor} contains a summation over both empty states and reciprocal lattice vectors (\textbf{G}), and the expression of the generalized dipole terms in Eq.~\eqref{yambo_dip} is such that matrix elements with large \textbf{G} are governed by high-energy KS states~\cite{van_setten_automation_2017}. Furthermore, the interdependence is non-trivial, given the presence of an inversion in Eq.~\eqref{chi_dyson} that further enters in the evaluation of the correlation self-energy, Eq.~\eqref{Sigma_c_yambo}. Given the lack of an efficient ``recipe''  to carry out this non-trivial, coupled convergence, one has to resort to an iterative convergence of each of the two parameters by fixing the other, in an alternating way. This procedure, in addition of being tedious, is computationally very expensive, sometimes representing the most cumbersome part of a GW calculation.

\subsection*{Bethe-Salpeter Equation}

Starting from the GW quasi-particles, the solution of the BSE~\cite{strinati1988application,onida_electronic_2002} can give access to optical properties of materials via the macroscopic dielectric function:
\begin{equation}
\begin{split}
    & \epsilon_M(\omega,\mathbf{q}) =  1 -  \frac{2}{V N_q }v(\mathbf{q})\sum_{\lambda}\Phi_{\lambda}(\mathbf{q}) \\
&    \times \left[ \frac{1}{E^{\lambda}(\mathbf{q}) - (\omega+i\eta)} + \frac{1}{E^{\lambda}(\mathbf{q}) + (\omega+i\eta)} \right],
    \label{eps_macro}
\end{split}
\end{equation}
where $V$ is the volume of the unit cell; $N_q$ is the number of $\textbf{q}$-points sampling the Brillouin zone (BZ); $E^{\lambda}(\mathbf{q})$ is the eigenvalue of the exciton $\lambda$ at momentum $\textbf{q} = \textbf{k}-\textbf{k}'$, and the corresponding exciton oscillator strength $\Phi_{\lambda}(\textbf{q})$ is defined as:
\begin{equation}
    \Phi_{\lambda}(\textbf{q})=\left|\sum_{(vc,\textbf{k})} \bra{v\textbf{k}-\textbf{q}} e^{-i\textbf{q}\cdot \textbf{r}} \ket{c\textbf{k}} A^{\lambda}_{vc\textbf{k}}(\textbf{q}) \right|^2.
    \label{eq:exc_oscill}
\end{equation}

Equation~\eqref{eq:exc_oscill} contains a summation over valence and conduction bands ($v,c$), and \textbf{k}-point mesh, which are the main parameters to converge for a BSE calculation. The terms $A^{\lambda}_{vc\textbf{k}}(\textbf{q})$ represent the weight of each electron-hole transition contributing the exciton $\lambda$, as resulting from the solution of the BSE. The summation over bands mainly defines the range of energies under investigation. The \textbf{k}-point mesh is connected to the accuracy with which we describe the exciton composition in terms of single-particle $v \rightarrow c$ transitions over the entire BZ, and it is usually significantly larger than the one needed to converge the corresponding GW band structure.
Additional convergence parameters are the number of \textbf{G} vectors used for expanding the KS wave-functions in the transition matrix elements, as defined in Eq.~\eqref{yambo_dip}, and for Fast-Fourier-Transform (FFT) operations, as well as the plane-wave expansion of the BSE kernel (both direct and exchange terms). These parameters are usually inherited from GW convergences.

\section{Results and discussion}

\subsection*{Description of the convergence surface}
\label{sec:conv}

The above-described coupled convergence of the parameters, combined with a much worse scaling than DFT and more computationally and memory demanding calculations, call for efficient procedures to describe and explore the convergence space in GW-BSE simulations. A possible strategy is to describe the convergence space in terms of an analytic function of the parameters~\cite{schindlmayr_analytic_2013, klimes_predictive_2014, maggio+gwsmall, van_setten_automation_2017, rangel_repr_2020,rasmussen_towards_2021}. 
For a general (N+1) dimensional space, a model convergence surface $f(\textbf{x})$ that represents the value of a given observable $E($\textbf{x}$)$ (e.g., quasiparticle energies or excitonic eigenvalues) as a function of the $N$ parameters $\textbf{x} = [x_1,...,x_N]$ can be defined as:
\begin{equation}
    \label{multi_over_x}
    f(\textbf{x}) = \prod_i^N \left( \frac{A_i}{x_i^{\alpha_i}} + b_i \right),
\end{equation}
where $A_i$, $b_i$ and $\alpha_i$ are free fitting parameters, and  $B = \prod_i^N b_i$ is the extrapolated converged value. Of course, the accuracy of the latter depends on the actual region of the parameter space explored for the evaluation of the convergence behaviour, i.e., how far is the extrapolated value from the exact one. As such, $B$ might not always be a good choice for guiding the search of the convergence parameters.

In the following, we introduce conditions on the mixed partial derivatives of Eq. \eqref{multi_over_x} as a way to address the parameter interdependence. Indeed, the adoption of an analytical form for the description of the convergence space has the clear advantage of enabling the calculation of all-order derivatives, once the fitting parameters are known. On the other hand, not taking directly into account this interdependence can result in a very tedious convergence procedure, as it would require the cyclic repetition of multiple univariate convergences, as mentioned in the Introduction (further details are provided in \suppinfo and Ref.~\onlinecite{yambo2}). 

For the expression in Eq. \eqref{multi_over_x}, the gradient components are:
\begin{equation}
    \label{f_prime}
    f_{x_i}'(\textbf{x}) =  -\alpha_i \frac{A_i}{x_i^{\alpha_i+1}} \prod_{j \neq i}^N \left( \frac{A_j}{x_j^{\alpha_j}} + b_j \right),
\end{equation}
while second derivatives are:
\begin{equation}
    f_{x_i}''(\textbf{x}) =  \alpha_i(\alpha_i+1) \frac{A_i}{x_i^{\alpha_i+2}} \prod_{j \neq i}^N \left( \frac{A_j}{x_j^{\alpha_j}} + b_j \right),
    \label{f_sec}
\end{equation}
\begin{equation}
    f_{x_i,x_j}''(\textbf{x}) =  \alpha_i \alpha_j \frac{A_iA_j}{x_i^{\alpha_i+1}x_j^{\alpha_j+1}} \prod_{k \neq i,j}^N  \left( \frac{A_k}{x_k^{\alpha_k}} + b_k \right).
    \label{f_sec_partial}
\end{equation}
The asymptotic region of the convergence surface can be determined by imposing, for each parameter $x_i, x_j$ with $i,j=1,...,N$, two conditions:
\begin{equation}
\label{derivatives_conditions}
        \left\{ \begin{array}{ll}
            |f_{x_i}'(\textbf{x})|< \Delta_i \\[7pt]
            |f_{x_i,x_j}''(\textbf{x})|<  \Delta_{ij}
        \end{array} \right.
\end{equation}
The first condition determines the region in which the convergence surface becomes flat (thus approaching convergence), whereas the condition on second partial derivatives ensures that the $N$ parameters are no longer interdependent. The threshold values $\Delta_i$ and $\Delta_{ij}$ can be tuned according to the desired accuracy.~\footnote{According to our experience, the asymptotic region can be safely identified by choosing $\Delta_i = 5 \cdot 10^{-5}$ and $\Delta_{ij} = 1\cdot 10^{-8}$.} Once this asymptotic region has been determined, a guess for the converged value of $f(\textbf{x})$, $E_{guess}$, is made, and its accuracy is then checked and validated according to the automated algorithm described below.

\subsection*{Convergence algorithm}
%
\begin{figure}[t]
    \centering
    \includegraphics[width=0.85\columnwidth]{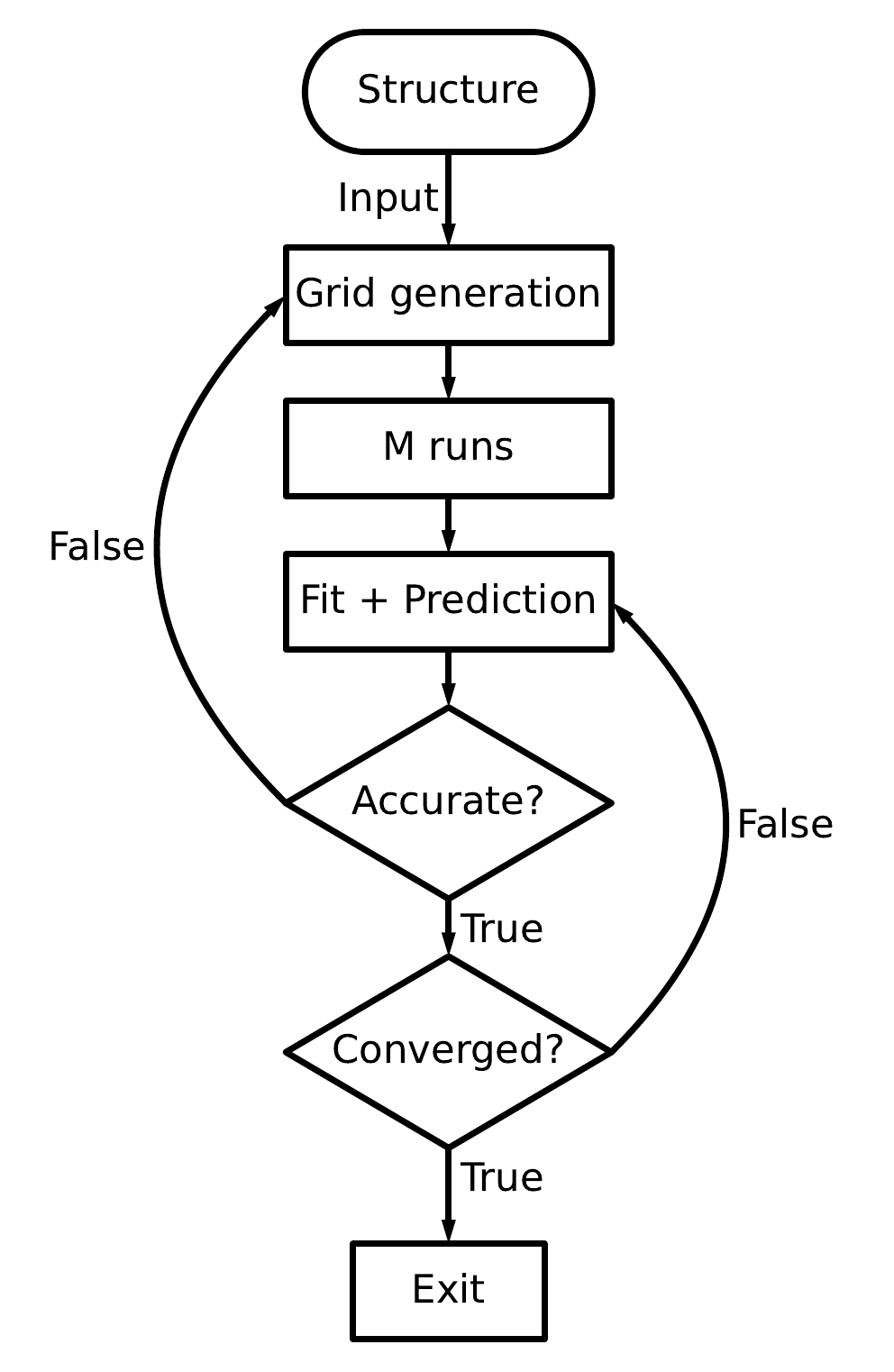}
    \caption{ \textbf{Flowchart of the convergence algorithm}. After generating the grid for the N-dimensional parameter space, a subset of M simulations is performed. The results are then fitted to predict the converged parameters. Finally, the accuracy (Eq.~\eqref{eq:acc_conv_conditions}a) and the convergence (Eq.~\eqref{eq:acc_conv_conditions}b) of the prediction are verified, and the procedure iterated, if needed.}
    \label{fig:schema_alg}
\end{figure}
%
Figure ~\ref{fig:schema_alg} schematically depicts our convergence algorithm, which is purposed to obtain the desired accuracy on GW-BSE results with the least possible number of calculations. In the following, we consider $N_b$ and $G_{\text{cut}}$ as the two interdependent parameters to be converged simultaneously. We remark that, while the algorithm is specifically designed to handle coupled convergences, it can be successfully used to accelerate convergence tests with respect to any other parameter, such as the BZ \textbf{k}-point mesh or the FFT grids.

The first step (i) consists in the construction of the N-dimensional space of parameters as a grid of equally spaced points, with spacing and ranges provided from input. 
It is worth noting that, to fit the functional form of Eq. \eqref{multi_over_x}, one needs to generate a grid with minimum 3$N$ points, since $f(\textbf{x})$ contains three free fitting parameters for each of the $N$ dimensions of the parameter space. However, a further reduction on the minimal grid size (that is, on the minimum number of calculations to perform) can be obtained by fixing the power-law dependence, $\alpha_i$, to a given value, as suggested in Ref.~\citenum{van_setten_automation_2017}, thus resulting in a minimal grid size of 2N.
Usually, much denser grids are generated, where (ii) M$_0 \geq$ 2$N$ calculations are performed on a subset of the grid points, chosen such as to efficiently sample the parameter space. 

Next (iii), the results of the calculations are fitted by using the expression Eq.~\eqref{multi_over_x}. As mentioned above, the power laws are fixed to given values: $\alpha_i \in$ $\{$1,2$\}$ $\forall$ i=1, ...N, and the one resulting in the lowest mean squared error is chosen.
The asymptotic region is then identified by computing the first and second order derivatives (Eqs.~\eqref{f_prime}-\eqref{f_sec_partial}), and imposing the conditions in Eq.~\eqref{derivatives_conditions}. Given the asymptotic region, a guess converged value is selected $E_{guess}$= $f(N^0_b, G^0_{cut})$, where ($N^0_b,G^0_{cut}$) are the lowest values that can be chosen for the parameters such that $E_{guess}$ is within a desired convergence threshold $\Delta$ with respect to the asymptotic region.

To establish the accuracy and convergence of the fitting procedure, (iv) $E(N^0_b, G^0_{cut})$ is computed and compared to the outcome of the fit $E_{guess}$, by considering the chosen threshold $\Delta$ (see Eq.~\eqref{eq:acc_conv_conditions}a below). If the accuracy condition is satisfied, we need to check the convergence of the fitting procedure, that is, a new pair of parameters $(N^1_b, G^1_{cut})$ is obtained from the fit by adding the $(N^0_b, G^0_{cut})$ point to the initial M$_0$ grid, and compared to the previous one (see Eq.~\eqref{eq:acc_conv_conditions}b below). The last step is repeated until convergence is reached. If the accuracy condition is not satisfied, the grid is instead shifted toward higher values of the parameters, and the steps (ii)-(iv) are repeated until the two conditions:
\begin{subequations}
    \begin{align}
\label{analytical_forms}
   &        |E(N^{j}_b, G^{j}_{cut}) - E^{fit}(N^{j}_b, G^{j}_{cut})|<\Delta \\[7pt]
    &        (N^{j+1}_b, G^{j+1}_{cut}) = (N^{j}_b, G^{j}_{cut}), 
   \end{align}
   \label{eq:acc_conv_conditions}
  \end{subequations}
are simultaneously satisfied for the $j$-th iteration. 

\subsection*{The \texttt{aiida-yambo} plugin and automated workflows}
The above convergence algorithm has been implemented in the new version of the \texttt{aiida-yambo} plugin~\cite{aiida-yambo-github}, which is meant to fully automate GW-BSE calculations by interfacing the \textsc{Yambo} project~\cite{yambo1,yambo2} and the AiiDA informatics infrastructure and workflow management system~\cite{huber_aiida_2020,uhrin_workflows_2021}. The automation concerns input generation, scheduler submission, and output parsing phases. \footnote{The output parsing of the  \texttt{aiida-yambo} plugin is partially done by using \textsc{yambopy} functions~\cite{yambo2}.} Thanks to the AiiDA infrastructure, links between single calculations are managed on the fly by ad-hoc, \textit{dynamic} workflows (the so-called \textit{workchains} in the AiiDA jargon), i.e. their execution path is not fixed, but can depend on the results of completed calculations. This allows for the implementation of complex logics, such as those characterizing the convergence algorithm and GW band interpolation that we propose in this work. Moreover, each calculation, together with inputs and outputs, is stored in the AiiDA relational database, thus ensuring data provenance and full reproducibility of results.

\begin{figure}[tbh!]
    \centering
    \includegraphics[width=0.95\columnwidth]{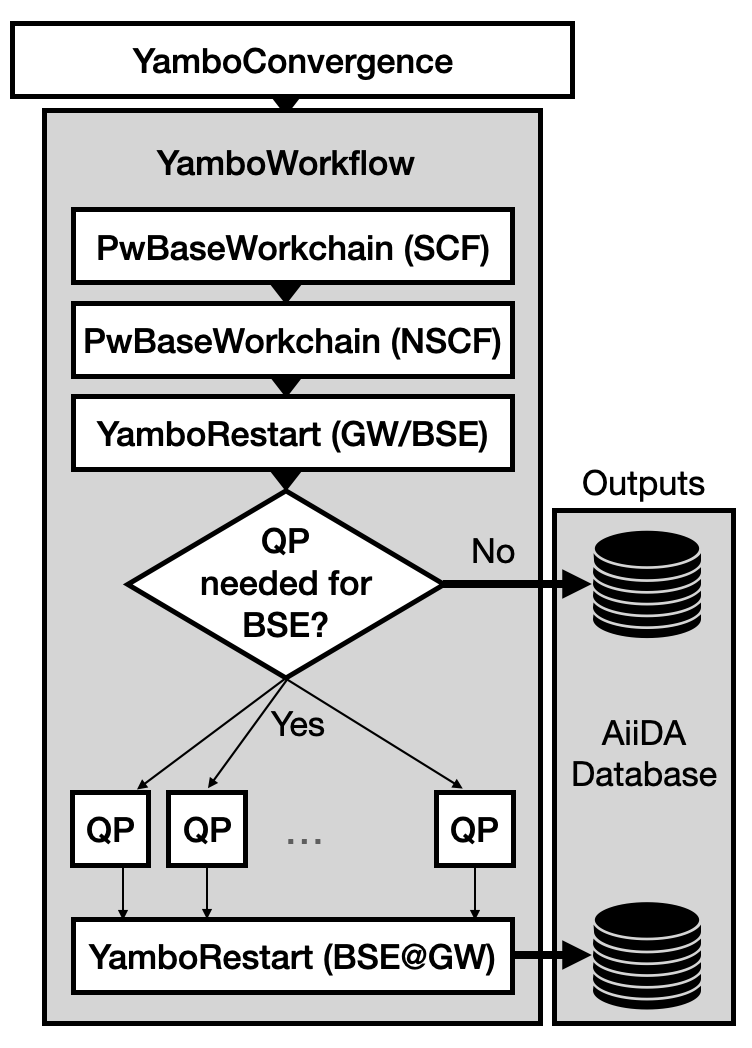}
    \caption{\textbf{Hierarchical structure of the \texttt{aiida-yambo} workchains}. The highest level workflow is \texttt{YamboConvergence}, which calls multiple \texttt{YamboWorkflow} workchains. \texttt{YamboWorkflow} comprises all the steps needed to perform individual GW-BSE calculations from scratch. In case of failures, it calls the \texttt{YamboRestart} workchain for automatic error handling. The outputs are stored in the AiiDA database in a human readable fashion, and are easily accessible and shareable by the user.}
    \label{fig:schema_plug}
\end{figure}

Currently, the \texttt{aiida-yambo} plugin supports quasiparticle ($G_0W_0$ and COSHEX~\cite{Hedin_1965} level) and optical properties (IP-RPA and BSE) simulations, as well as interfaces with different codes (e.g., \textsc{Quantum ESPRESSO} and \textsc{Wannier90}). 
These options are implemented in the \texttt{YamboCalculation} and \texttt{YppCalculation} classes, which manage individual simulations (including data interfacing) that can be performed by using the \textsc{Yambo} code.
On top of them, task-specific workflows are implemented, and organized in a modular way, in order to automate tasks of increasing complexity. 
In particular, the \texttt{aiida-yambo} plugin contains three main workflows, each of them targeting a precise task:
\begin{itemize}
    \item \texttt{YamboRestart}: automation of error handling and restart for each \texttt{YamboCalculation};
    \item \texttt{YamboWorkflow}: automation of the single GW or BSE flow (composed of several interlinked steps, explained in the following);
    \item \texttt{YamboConvergence}: automation of the convergence (composed of multiple \texttt{YamboWorkflow} runs).
\end{itemize}
Their nested organization is shown in Fig.~\ref{fig:schema_plug}.
%
%

The highest level workflow is represented by the \texttt{YamboConvergence} workchain, which implements the full automation of the convergence algorithm of Sec.~\ref{sec:conv}, thus allowing for all \textsc{Yambo} simulations to be organized on the fly, without any external user intervention. 
The user is only requested to provide a python list containing the information on the parameter space to be explored. An example of such input reads:
\begin{lstlisting}
[
    {
        'var': [
            'BndsRnXp', 
            'GbndRnge', 
            'NGsBlkXp'
            ],
        'start': [50, 50, 2],
        'stop': [400, 400, 10],
        'delta': [50, 50, 2],
        'max': [1000, 1000, 36],
        'what': ['gap_GG'],
        'conv_thr': 0.1,
        'conv_thr_units': 'eV',
        },
]
\end{lstlisting}
%
where the \textsc{Yambo} variables “BndsRnXp" and “GbndRnge" govern the convergence over empty states, $N_b$,  to be carried out jointly with that on the size of the response matrix, $G_{\text{cut}}$ (“NGsBlkXp" variable). The edges of the grid (“start" and “stop") and its spacing (“delta"), together with an upper bound of the parameter space (“max"), limiting the search to computationally accessible calculations, are also set by the user. The key “what" indicates the quantity to be converged -- in our example, the direct band gap of the material at $\Gamma$ point -- up to a given convergence threshold $\Delta$ (“conv$\_$thr" key). The output summarizes the convergence history and allows the user to easily parse the converged simulation. \texttt{YamboConvergence} allows one to converge several many-body quantities, like quasiparticle levels,  band-gaps, as well as optical excitation energies. Notably, the convergence block in Fig.~\ref{fig:schema_plug} can be skipped if converged parameters are already known.

Each single GW (BSE) calculation is instead automated within \texttt{YamboWorkflow}, which is the core workchain of the plugin that takes care of performing all the steps needed in a typical \textsc{Yambo} simulation -- from preliminary self-consistent (SCF) and non-self-consistent (NSCF) DFT calculations to the actual GW (BSE) calculations, and the related post-processing. The workflow ensures a robust interoperability between DFT and MBPT codes (\textsc{Quantum ESPRESSO} and \textsc{Yambo}, respectively), and links subsequent calculations, interfacing the data automatically. In practice, \texttt{YamboWorkflow} encodes the specific flowchart underlying each requested calculation, and allows for its dynamic execution according to the instructions provided in input. This implies performing all the intermediate steps needed for a specific calculations without the need of instructing them explicitly, or, on the contrary, to skip some of the intermediate steps for which parent calculations are available, fully exploiting the \texttt{YamboWorkflow} provenance information.

To support a restart mechanism in case of code failures, \texttt{YamboWorkflow} takes advantage of the \texttt{YamboRestart} workchain, a sub-level workflow that encodes an automatic error handler (inherited from the AiiDA \texttt{BaseRestartWorkchain} class) which, depending on the encountered failure, automatically instructs a restart run.
For out-of-memory errors or failures connected with insufficient wall-time requests, \texttt{YamboRestart} automatically resubmits the calculation by appropriately changing the requested resources (e.g., the maximum wall-time and the MPI/OpenMP balance); parallelization errors are managed by overwriting the parallelism variables set in input by the user with the default parallelism decided on the fly by \texttt{yambo}. 
In all these cases, an efficient, CPU-time saving restart mechanism is implemented, which avoids to restart unfinished runs from scratch by automatically retrieving and enabling the reuse of stored data files.

As a final issue, we would like to discuss the possibility to develop protocols for MBPT calculations. 
Indeed, most of the DFT-based AiiDA plugins enable the use of protocols~\cite{huber_common_2021}, that is, the possibility of creating inputs with pre-populated default values for several parameters. Such protocols are usually code-agnostic and robust, given the high level of reproducibility of DFT with different quantum engines~\cite{lejaeghere_reproducibility_2016}. Moreover, their reliability is guaranteed by means of large scale studies spanning systems with a wide variety of characteristics (i.e., metals, semiconductors, dimensionality and so on)~\cite{prandini_precision_2018}. 

Concerning MBPT calculations, the possibility to define protocols is still an open issue~\cite{van_setten_automation_2017}. First of all, code-agnostic parameters are not at all easy to be determined as it is for DFT-based codes, because the MBPT implementations and the subsequent definition of parameters can differ in very many aspects, as highlighted in the Introduction section. Secondly, the high computational cost of these simulations has limited so far the number of systems to be studied extensively, which is crucial to define a reliable statistics on convergence parameters. Last but not least, DFT-based protocols usually result in safe but overconverged parameters, an approach that might lead to unfeasible calculations when moving to the GW-BSE framework. 

For all these reasons, we believe that an efficient, fully automated convergence tool, as the one presented here, is currently the most valuable solution. Nonetheless, also in view of possible future developments, the \texttt{aiida-yambo} plugin provides an implementation for a protocol framework for both GW and BSE simulations, which is currently pre-populated on the basis of previous experience on a limited subset of systems. Such protocols concern several parameters connected to the main \texttt{yambo} input variables, such as the FFT grids (“FFTGvecs"), the summation over empty states (“BndsRnXp" and “GbndRnge"), the plane-wave expansion for the polarizability (“NGsBlkXp") and the BZ \textbf{k}-point sampling. A detailed documentation on these and other aspects concerning the \texttt{aiida-yambo} plugin is provided elsewhere~\cite{aiida-yambo-docs}. 
\subsection*{Automatic GW bands interpolation: the \texttt{aiida-yambo-wannier90} plugin}
%
%
\begin{figure}[tbh!]
  \includegraphics[width=0.7\columnwidth]{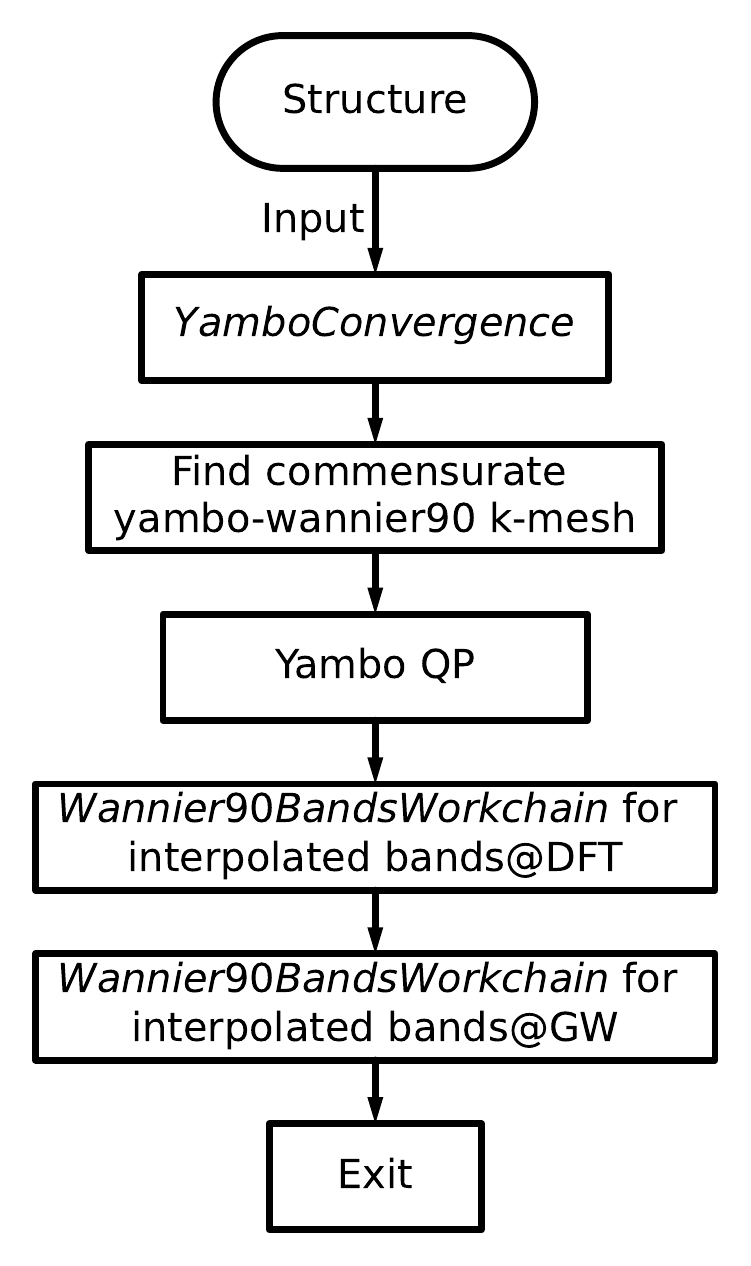}
  \caption{\textbf{Flowchart of the \texttt{YamboWannier90WorkChain}} for the
    Wannier interpolation of GW band structures.
    After performing the \texttt{Yambo} convergence, the workflow searches for a commensurate k-point meshes for \texttt{Yambo} and \texttt{Wannier90}, and carries out the corresponding Yambo QP calculation. Given the QP corrections, the workflow proceeds with the Wannierization and the band interpolation at DFT and GW level.} 
 \label{fig:gww90_flowchart}
\end{figure}
%

GW band interpolation from Wannierization is a crucial task in order to obtain
the most accurate quasiparticle band structure with the lowest computational cost. This task is encoded in the \texttt{aiida-yambo-wannier90} plugin~\cite{aiida-yambo-wannier90-github,aiida-yambo-wannier90-docs}.
Essentially, the plugin provides a meta-workflow, called
\texttt{YamboWannier90WorkChain}, which utilizes the automation and error
handling of the underlying \texttt{aiida-yambo} and
\texttt{aiida-wannier90-workflows} plugins for GW convergence and
Wannierization, respectively. The flowchart of the workflow is summarized in
Fig. \ref{fig:gww90_flowchart}. 
Starting from a given crystal structure, the workflow first launches a \texttt{YamboConvergence} workflow for automatic
convergence. Then, it finds the minimal commensurate mesh with the \texttt{Wannier90} ones that satisfies the GW convergence conditions (see below).
Thirdly, it runs \texttt{YamboWorkflow} to compute all the quasiparticle corrections required for the Wannierization on the commensurate mesh, and a subsequent \texttt{ypp} calculation (by means of \texttt{YppRestart}) to extract the GW corrections in a \texttt{Wannier90} \texttt{eig} file format. 
Fourthly, the workflow Wannierizes the KS wavefunctions, saving the unitary transformation matrices of maximal localization, and interpolates the band structure. Finally, the workflow performs the Wannierization procedure at $G_0W_0$ level, which consists in incorporating the GW corrections into the DFT eigenvalues,
and interpolating the band structure by using the DFT Wannierization outcomes.

\begin{figure}
  \includegraphics[width=\linewidth]{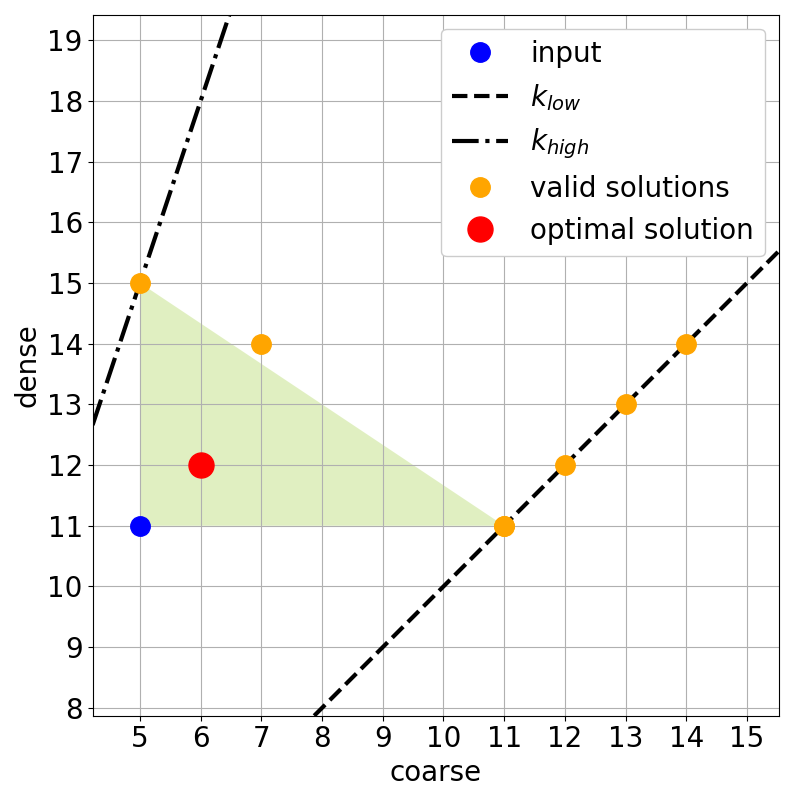}
  \cprotect\caption{\textbf{Recipe to find commensurate meshes for GW and \texttt{Wannier90} calculations}.
    Using input meshes (11, 5) as an example, final commensurate meshes (12,6) are found.
    The $k_{high}$ and $k_{low}$ lines, that intersect in the origin, are respectively the imposed upper and lower bound for
    searching the commensurate meshes; the orange dots are possible solutions; the red dot is the chosen solution, which is the closest to the input in the metric of $\ell^{1}$ norm.
  } \label{fig:commensurate_mesh}
\end{figure}

A crucial step of the workflow is finding a commensurate mesh for both
GW QP calculations and Wannierization. Indeed, the GW mesh resulting from automated convergence might not always be compatible with the mesh required by \texttt{Wannier90} to ensure interpolation accuracy.
Notably, considering a Monkhorst-Pack (MP) grid for the Wannierization, the corresponding GW mesh must be an integer multiple of the MP grid.
We here propose a recipe to find the minimal commensurate meshes for GW and \texttt{Wannier90} calculations, as depicted in Fig.~\ref{fig:commensurate_mesh}.
Considering $n_d$ as the number of k-points chosen by the \texttt{YamboConvergence} workflow, and $n_c$ the number of k-points chosen by the
Wannierization protocol (typically based on a k-point spacing, 0.2 \AA$^{-1}$)~\cite{Vitale2020} , the
target is to find a new ($n_d^\prime$, $n_c^\prime$) such that the dense mesh
$n_d^\prime = k \cdot n_c^\prime$, where $k \in \mathbb{N}$, i.e., natural number. The given input $(n_d, n_c)$ restricts the search space to a sector bounded by
$k_{\text{low}}$ and $k_{\text{high}}$ (see Fig.~\ref{fig:commensurate_mesh}),
where $k_{\text{low}} = 1$ ($n_d^\prime = n_c^\prime = \max(n_d, n_c)$ is always a good solution), and $k_{\text{high}} = \lceil\frac{n_d}{n_c}\rceil$, where $\lceil\cdot\rceil$ indicates the ceiling integer. 
The search always succeeds since $s_{\text{low}} = (\max(n_d, n_c), \max(n_d, n_c))$ and
$s_{\text{high}} = (k_{\text{high}} \cdot n_c, n_c)$ are already two good
solutions. In fact, the optimal solution is often inside the triangular region
determined by the input $(n_d, n_c)$, $s_{\text{low}}$, and $s_{\text{high}}$. 
The final solution is chosen according to the $\ell^1$
distance to the input, such to ensure the minimal increase in computational cost. 
It is also possible to change the metric, e.g., pushing the solution towards
increasing the Wannier mesh or GW mesh, depending on which calculation is
less cumbersome. The aforementioned recipe is repeated for each of the three dimensions
of the MP grid.

%
\subsection*{Validation of the workflows}
\label{sec:results}
%
\begin{table*}[th!]
\begin{tabular}{ l|ccccccccccc } 
\hline
\hline
 & \\[-7pt]
{\bf System} & E$^{G_0W_0}_{gap}$ (eV) & E$^{ref}_{gap}$ (eV) & $N_b$ & $G_{\text{cut}}$ (Ry) & $\rho_{\textbf{k}}$(\AA$^{-1}$) & $\Delta^{\Gamma}$(meV) & $\Delta^{\Gamma}_{\%}$\\[3pt]
\hline
 & \\[-7pt]
Si & 1.18 & 1.16~\cite{rangel_repr_2020} & 400 & 16 & 0.33 & 10 & 0.3\\ 
Diamond & 5.82 & 5.63~\cite{gao_2015} & 300 & 20 & 1 & 70 & 0.1\\ 
ZnO & 2.36 & 2.35~\cite{stankovski_g_2011} & 800 & 28 & 0.25 & 10 & 0.4\\ 
TiO$_2$ & 3.20 & 3.20~\cite{rangel_repr_2020} & 600 & 12 & 0.2 & 10 & 0.3\\ 
MoS$_2$ & 2.54 & 2.54~\cite{Rasmussen_Q0} & 400 & 8 & 0.25 & 10 & 0.3\\ 
hBN bulk & 6.28 & 6.30~\cite{huser_quasiparticle_2013} & 800 & 20 & 0.25 & 82 & 1.0\\ 
hBN 2D & 6.84 & 7.06~\cite{Rasmussen_Q0} & 1200 & 28 & 0.2 & 42 & 0.5\\[3pt]
\hline
\hline
\end{tabular}
\caption{\textbf{$G_0W_0$ convergence tests on prototypical semiconductors}. For each system, the minimum band gap $E^{G_0W_0}_{gap}$ (second column) is computed by using the parameters obtained by the automated convergence algorithm implemented in the \texttt{YamboConvergence} workchain, and compared with previous $GW$ results (third column). The considered parameters are the number of empty states, $N_b$, the energy cutoff on the \textbf{G} vectors, $G_{\text{cut}}$, and the irreducible Brillouin Zone (iBZ) \textbf{k}-points density $\rho_{\textbf{k}}$, expressed as the maximum distance between adjacent points along a reciprocal axis. The first two parameters are converged jointly. The last two columns include the convergence thresholds imposed on the G$_0$W$_0$ band gap at the $\Gamma$ point, both in absolute ($\Delta^{\Gamma}$) and relative ($\Delta^{\Gamma}_{\%}$) terms. }
\label{summary}
\end{table*}

The proposed convergence algorithm, as implemented in the \texttt{YamboConvergence} workchain, has been validated by performing convergence studies for the quasiparticle $G_0W_0$ gap of a set of prototypical semiconductors: silicon, diamond, ZnO, rutile TiO$_2$, monolayer MoS$_2$, bulk and monolayer hBN. 
The convergence addresses the direct band gap at the $\Gamma$ point with respect to the two coupled parameters, $N_b$ and $G_{\text{cut}}$, and the $\mathbf{k}$-point grid (in terms of \textbf{k}-points density $\rho_{\textbf{k}}$).
Once convergence is achieved, the minimum band gap $E^{G_0W_0}_{gap}$ is also computed. The results are summarized in Table~\ref{summary}, where, together with $E^{G_0W_0}_{gap}$, we report the final parameters resulting from the convergence procedure ($N_b$, $G_{\text{cut}}$, and $\rho_{\textbf{k}}$) as well as the convergence threshold (absolute, $\Delta^{\Gamma}$, and relative, $\Delta^{\Gamma}_\%$) adopted in each case. Our results are found in good agreement with previous findings, which are also reported in the Table~\ref{summary}. Deviations with respect to reference results can be ascribed to different GW implementations and/or different DFT starting points used in the related works~\cite{rangel_repr_2020}. 
Further details are contained in the \suppinfo. 

\begin{figure*}
    \centering
    \includegraphics[width=1.0\columnwidth]{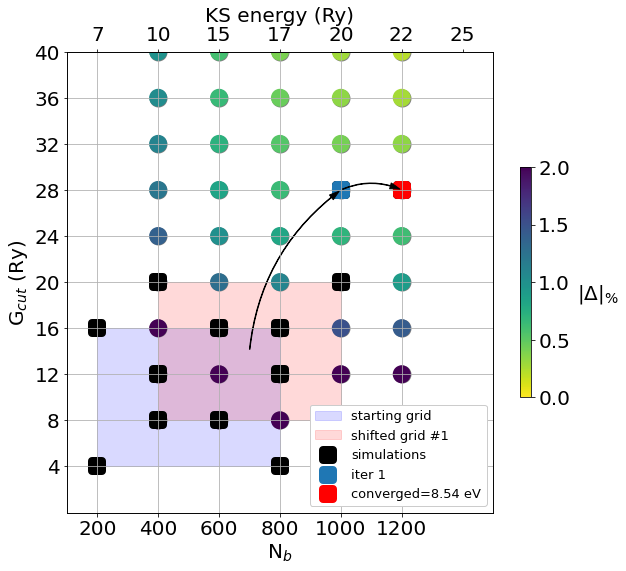}
    \centering
    \includegraphics[width=0.9\columnwidth]{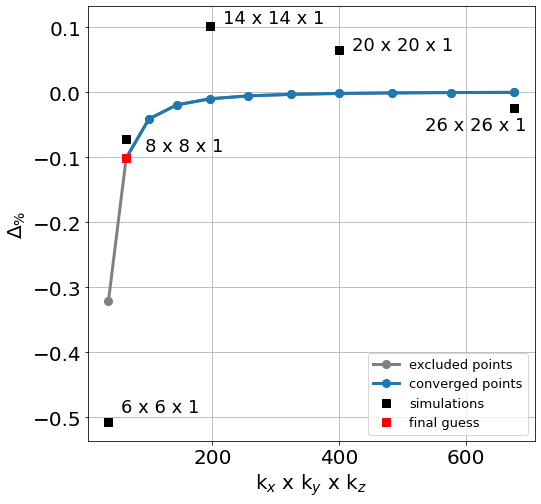}
    \caption{\textbf{Convergence algorithm applied to monolayer hBN}. 
    \textbf{(a)} Convergence of the direct quasiparticle band gap at the $\Gamma$ point with respect to the coupled parameters $N_b$ and $G_{\text{cut}}$. The blue shaded area represents the starting grid, while the red shaded area the shifted grid obtained after the first iteration, according to the flowchart in Fig.~\ref{fig:schema_alg}. Black squares represent the actual calculations performed by the workflow; the blue square is the first guess for the converged parameters; the red square indicates the final, converged point. The colormap specifies the relative error with respect to the converged point. A maximum absolute error of $\Delta$ = 42 meV is achieved for ($N_b$, $G_{\text{cut}}$) = (1200,28 Ry), corresponding to a maximum relative error of $\Delta_{\%}$ = 0.5 $\%$. \textbf{(b)} Convergence of the direct quasiparticle band gap at the $\Gamma$ point with respect to the \textbf{k}-mesh. The black points represent the actual calculations performed by the workflow, whereas the blue points are the ones obtained within the fitting procedure and used to predict the convergence. The final, converged mesh (red square) is achieved with five simulations.}
    \label{fig:hBN_2d}
\end{figure*}

Figure~\ref{fig:hBN_2d} shows the convergence procedure for monolayer hBN, considering both the joint convergence with respect to N$_b $ and $G_{\text{cut}}$ (panel a), and the single-parameter convergence with respect to the $\textbf{k}$-mesh (panel b). Starting from an input grid with N$_b \in$ [200,800] and G$_{cut} \in$ [4,16] Ry (Figure~\ref{fig:hBN_2d}a, blue shaded area), a subset of 6 calculations is performed (black squares). Since the converged guess is above the upper bound of the parameter space (see “max” variable in the plugin description), a new shifted grid (orange shaded area) is considered, and a first guess for the converged parameters is found from the fitting procedure (blue square), now satisfying the accuracy condition (Eq.~\eqref{eq:acc_conv_conditions}a). A new fit is performed including this additional point, which results in a new converged guess (red square). The procedure is repeated till the converged result is verified to be consistent with the prediction within the given threshold (Eq.~\eqref{eq:acc_conv_conditions}a) and to be the true converged point (Eq.~\eqref{eq:acc_conv_conditions}b). A similar path is followed for the $\textbf{k}$-mesh convergence (Figure~\ref{fig:hBN_2d}b): a limited number of calculations is initially performed (black squares), from which the fitting is evaluated (blue curve), and the smallest grid compatible with the given threshold is finally selected (here 8$\times$8$\times$1). We note that, despite the results of the simulations seem to have an oscillating behaviour with respect to the fitted curve, the error bar considered here (from the 14$\times$14$\times$1 mesh) is $\sim0.13\%$ of the band gap at $\Gamma$, i.e. $\sim10$ meV, in line with the accuracy of state-of-the-art GW calculations.

The dependence of the outcome of the algorithm on the input settings (e.g., the initial value of the parameters and the boundaries of the convergence space) is a key issue for evaluating the robustness of the algorithm itself. Indeed, the convergence procedure has been tested on the starting grid for monolayer MoS$_2$. By using two different grids, N$_b  \in$ [200,800], G$_{cut} \in$ [4,20] Ry and N$_b  \in$ [200,1200], G$_{cut} \in$ [8,24] Ry, the convergence point obtained from the workflow remains the same, i.e., (N$_b $,$G_{\text{cut}}$) = (400,8). Another important issue to evaluate is the efficiency of the algorithm. We notice that, in the case of the 2D-hBN convergence shown in the left panel of Figure~\ref{fig:hBN_2d}, only 14 calculations where required to reach convergence. Older implementations~\cite{yambo2} would have required $\geq$ 25 simulations to achieve a final result, with over 40\% reduction.

\begin{figure*}
    \centering
    \includegraphics[width=1\columnwidth]{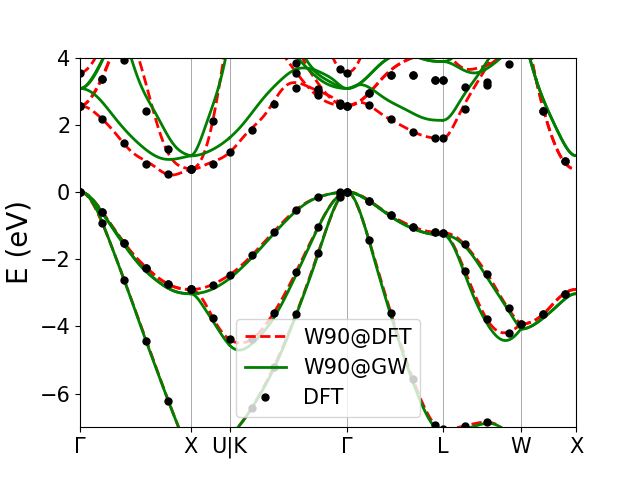}
    \includegraphics[width=1\columnwidth]{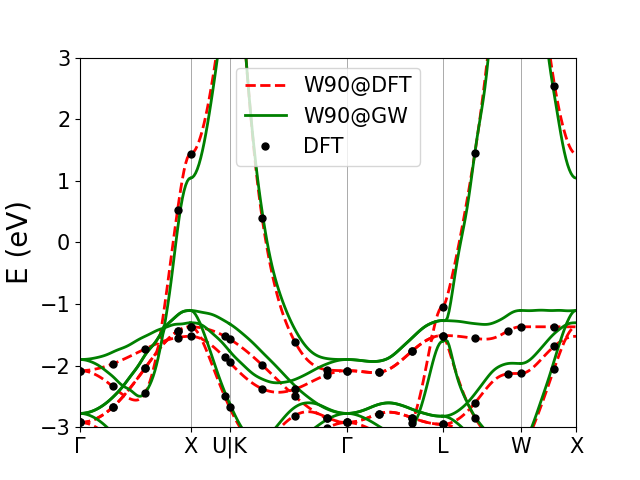}
    \caption{\textbf{Wannier interpolation of GW band structures}. Band structure of Si (a) and Copper (b). Interpolated bands are plotted for both DFT (red dashed line) and GW (green solid line) eigenvalues, as compared to the DFT computed bands (black dots).} 
    \label{fig:Si_GWW90}
\end{figure*}

Next, the \texttt{YamboWannier90WorkChain} as been tested on bulk Si and Cu, in order to validate  the automatic Wannier interpolation of GW band structures for both semiconductors and metals. Results are plotted in Fig. \ref{fig:Si_GWW90}, where we compare the DFT bands, the Wannier interpolated DFT, and the Wannier interpolated GW bands. For Si (Fig. \ref{fig:Si_GWW90}a), the comparison between computed and interpolated DFT bands shows that the results are almost identical, indicating the accuracy of the Wannierization of the KS wavefunctions. Moreover, the typical band gap opening upon inclusion of GW corrections is found when comparing KS and QP band structures. 

For Cu (Fig. \ref{fig:Si_GWW90}b), we obtain a discrepancy of $\sim$ 10 meV around the Fermi energy (here set to zero) between computed and interpolated bands at the DFT level. Better accuracy can be achieved imposing more stringent values of the involved parameters. At QP level, the GW correction is very small around the Fermi level ($\sim$ 37 meV), but still not negligible. Here, the GW convergence is more stringent than for Si, especially concerning the \textbf{k}-point mesh. Indeed, denser grids are needed to account for the contribution of intra-band transitions in the \textbf{q}$\rightarrow$0 limit, which is crucial for metallic systems but not explicitly included within the plasmon pole approximation~\cite{hybertsen_electron_1986,godby_metal-insulator_1989}. 
Considering the converged parameters, ($N_b$, $G_{\text{cut}}$, $\rho_{\textbf{k}}$) = (400, 18 Ry, 0.2 \AA$^{-1}$), the quasiparticle evaluations required to interpolate the bands for the minimum converged \textsc{Wannier90} \textbf{k}-point mesh (16$\times$16$\times$16) become 2900. This quite large number of QP can be easily computed using the \texttt{YamboWorkflow} workchain thanks to the possibility to split the QP calculation in several runs, each of them computing only a fraction of the GW corrections, and then collecting all the data in a final database well-suited for the \texttt{YamboWannier90WorkChain}. Since the number of QP corrections to compute can be quite high, in the Supporting Information (section S.IIIA)
we suggest an effective way to reduce the number of the required calculations for cases when accurate QP corrections are needed only in a limited energy region, e.g. around the Fermi energy, and energies outside of the chosen region can be approximated e.g. through a scissor and stretching correction.

%
%
\section{Conclusions}
In this work, we have presented the successful design and implementation of advanced algorithms in state-of-the-art GW (BSE) calculations, that is, convergence between interdependent parameters, error handling and automatic band interpolation by means of Wannierization. We validated the tools on selected cases among semiconductors and metallic systems.
The results contained in this work clearly show the power of these
newly developed workflows for the automated study of excited states properties
of materials, paving the way for achieving high-throughput MBPT studies. Thanks to these developments and within the next-generation of pre-exascale and exascale supercomputers, these simulations may become extensively and routinely performed by the materials-science community in the near future.
%
\section{Methods}

For all the systems studied here, we used symmetrized geometries in such a way to reduce the computational cost of simulations. We do not expect relevant differences in the results obtained with fully-relaxed structures. DFT simulations were carried out by using the \textsc{Quantum ESPRESSO} simulation package, which implements plane-wave basis set and pseudopotential approach. The KS-DFT exchange-correlation functional was approximated using GGA-PBE~\cite{Perdew-Burke-Ernzerhof1996PRL}, through the optimized norm-conserving Vanderbilt (ONCV) SG15~\cite{oncv1,oncv2} pseudopotentials. In the case of ZnO, we adopted Local Density Approximation (LDA), to compare the results with the existing literature~\cite{stankovski_g_2011,rangel_repr_2020}, and PseudoDojo pseudopotentials~\cite{van_setten_pseudodojo_2018}. GW and BSE results were obtained by means of the \textsc{Yambo} code. The frequency dependence of the screened interaction potential was approximated by  using the Godby-Needs plasmon pole approximation~\cite{godby_metal-insulator_1989} (GNPPA), and the quasiparticle energies were calculated according to the G$_0$W$_0$ approximation~\cite{Hedin_1965,onida_electronic_2002}, as implemented in \textsc{Yambo}. 
The Bruneval-Gonze technique~\cite{BG} was used to reduce the number of empty states $N_b$ needed for the construction of the correlation Self-Energy $\Sigma_c$ (Eq.~\eqref{Sigma_c_yambo}).
For low-dimensional systems, spurious interactions between supercell replica were avoided using a slab truncation of the Coulomb potential~\cite{rozzi_exact_2006} along the non-periodic direction; its divergences are cured by means of the Random Integration Method~\cite{pulci_ab_1998} (RIM), which also accelerates convergence with respect to the BZ sampling. For 2D systems, specifically, we adopted a recently developed accelerating technique based on stochastic integration of the screened potential~\cite{guand_RIMW}, which allows to have GW-converged results using reduced Monkhorst-Pack \textbf{k}-points grids, just slightly denser than the DFT one.
Finally, Wannierization and band interpolations are performed by means of the \textsc{Wannier90} code.
All the simulations are performed using the automated workflows implemented in the \texttt{aiida-yambo} and \texttt{aiida-yambo-wannier90} plugins, developed for the AiiDA platform and presented here as part of the results achieved in this work. Input parameters are generated using the protocols procedure, as implemented in the corresponding plugins.
\section*{Data availability}
The data supporting the findings of this paper are available on the Materials Cloud~\cite{talirz_materials_2020} at \href{https://doi.org/10.24435/materialscloud:6w-qh}{https://doi.org/10.24435/materialscloud:6w-qh}. Results obtained in this work can be reproduced by means of the example scripts delivered within the \texttt{aiida-yambo} and \texttt{aiida-yambo-wannier90} plugins. 

\section*{Code availability}
All the codes used in this work are fully available to the community by means of their repositories, and supported by appropriate documentations. The \textsc{Yambo} code is accessible at \href{https://www.yambo-code.eu/download}{https://www.yambo-code.eu/download/}. The \textsc{Quantum ESPRESSO} and \textsc{Wannier90} codes can be found, respectively, at \href{https://www.quantum-espresso.org/download}{https://www.quantum-espresso.org/download} and \href{http://www.wannier.org/download}{http://www.wannier.org/download}. 

The AiiDA infrastructure is available at \href{http://www.aiida.net/download}{http://www.
aiida.net/download}. AiiDA plugins can be downloaded from the corresponding GitHub repositories, already referenced throughout the manuscript, see Refs. \citenum{aiida-yambo-github,aiida-yambo-wannier90-github}.

\section*{Acknowledgements}
%
%
We acknowledge stimulating discussions with Nicola Marzari, Michael O. Atambo, Gianluca Prandini, Dario A. L. Valido, Alberto Guandalini and Fulvio Paleari.
This work was supported by: MaX -- MAterials design at the eXascale -- a European Centre of Excellence funded by the European Union's program H2020-INFRAEDI-2018-1 (Grant No. 824143); the European Union's Horizon 2020 research and innovation programme (Grant No. 957189, BIG-MAP, also part of the BATTERY 2030+ initiative, Grant No. 957213); SUPER (Supercomputing Unified Platform – Emilia-Romagna) from Emilia-Romagna PORFESR 2014-2020 regional funds; the Swiss National Science Foundation (SNSF) Project Funding (Grant No. 200021E$\_$206190 “FISH4DIET"); NCCR MARVEL, a National Centre of Competence in Research, funded by the Swiss National Science Foundation (Grant No. 205602). Computational time on the Marconi100 and Galileo100 machines at CINECA was provided by the Italian ISCRA program.
\section*{Author contributions}
M.B., J.Q., A.M. and N.S. contributed to the development of the \texttt{aiida-yambo} plugin; M.B. designed, implemented and tested the automatic convergence algorithm, and the other workflows belonging to the \texttt{aiida-yambo} plugin. 
J.Q. and M.B. implemented the \texttt{aiida-yambo-wannier90} plugin.
E.M., A.F., D.V., D.P, G.P. were responsible for the project supervision and coordination. M.B., J.Q., and D.P. wrote the manuscript with contributions from all authors.

The Authors declare no Competing Financial or Non-Financial Interests

\bibliographystyle{my_aip}
\bibliography{./bibliography,./codes, ./mbpt_theory,./suppinfo} 
%
\end{document}


\title{Supplementary Material for: Towards high-throughput many-body perturbation theory: efficient algorithms and automated workflows}

\author{Miki Bonacci}
\email[corresponding author:]{miki.bonacci@nano.cnr.it}
\address{FIM Department, University of Modena and Reggio Emilia, Via Campi 213/a, Modena, Italy}
\address{S3 Center, Istituto Nanoscienze, CNR, Via Campi 213/a, Modena, Italy}
\author{Junfeng Qiao}
\address{Theory and Simulation of Materials (THEOS) and National Centre for Computational Design and Discovery of Novel Materials (MARVEL), \'Ecole Polytechnique F\'ed\'erale de Lausanne, CH-1015 Lausanne, Switzerland}
\author{Nicola Spallanzani}
\address{S3 Center, Istituto Nanoscienze, CNR, Via Campi 213/a, Modena, Italy}
\author{Antimo Marrazzo}
\address{Dipartimento di Fisica, Universit\`a di Trieste, I-34151 Trieste, Italy}
\author{Giovanni Pizzi}
\address{Theory and Simulation of Materials (THEOS) and National Centre for Computational Design and Discovery of Novel Materials (MARVEL), \'Ecole Polytechnique F\'ed\'erale de Lausanne, CH-1015 Lausanne, Switzerland}
\address{Laboratory for Materials Simulations (LMS), Paul Scherrer Institut (PSI), CH-5232 Villigen PSI, Switzerland}
\author{Elisa Molinari}
\address{FIM Department, University of Modena and Reggio Emilia, Via Campi 213/a, Modena, Italy}
\address{S3 Center, Istituto Nanoscienze, CNR, Via Campi 213/a, Modena, Italy}
\author{Daniele Varsano}
\address{S3 Center, Istituto Nanoscienze, CNR, Via Campi 213/a, Modena, Italy}
\author{Andrea Ferretti}
\address{S3 Center, Istituto Nanoscienze, CNR, Via Campi 213/a, Modena, Italy}
\author{Deborah Prezzi}
\address{S3 Center, Istituto Nanoscienze, CNR, Via Campi 213/a, Modena, Italy}

\maketitle

\section{Convergence plots for the systems studied in this work}
In the following, we show the convergence plot for the systems studied in this work, as done for in the main text for the monolayer hBN case.  We converged the $\Gamma-\Gamma$ band gap with respect to the two coupled parameters $N_b$ and $G_{cut}$ and the $\mathbf{k}$-point grid as well, except for TiO$_2$ and ZnO. For diamond, we converged also the FFTGvecs parameter, governing the Fast-Fourier-Transform (FFT) size for the evaluation of the generalized dipole matrix elements.
\begin{figure}[h]
    \centering
    \includegraphics[scale=0.4]{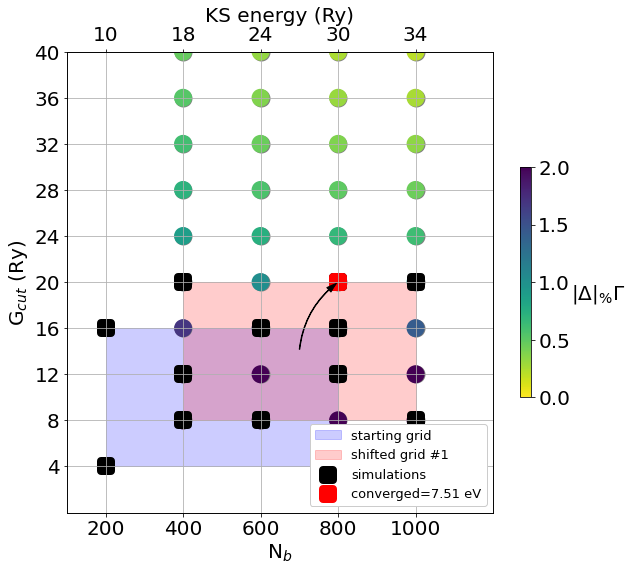}
    \caption{Convergence of $N_b$ and $G_{cut}$ for bulk hBN.}
\end{figure}
\begin{figure}[h]
    \centering
    \includegraphics[scale=0.4]{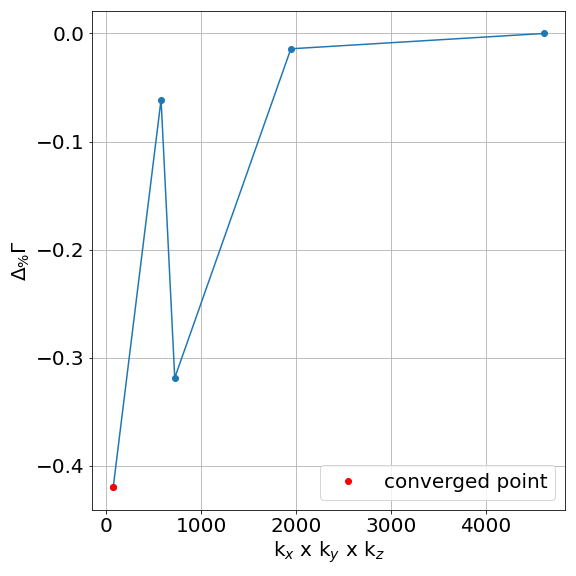}
    \caption{\textbf{k}-mesh convergence for bulk hBN.}
    \label{fig:hBN_2d_SI}
\end{figure}
\begin{figure}[h]
    \centering
    \includegraphics[scale=0.4]{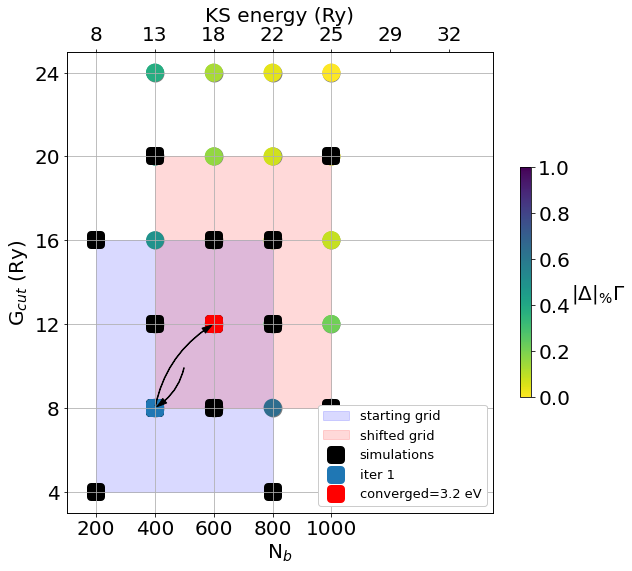}
    \caption{Convergence of $N_b$ and $G_{cut}$ for Rutile TiO$_2$.}
\end{figure}
\begin{figure}[h]
    \centering
    \includegraphics[scale=0.4]{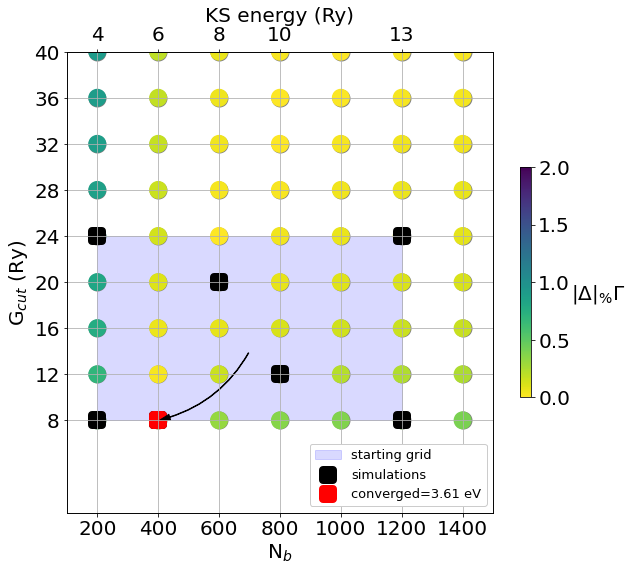}
    \caption{Convergence of $N_b$ and $G_{cut}$ for monolayer MoS$_2$.}
\end{figure}
\begin{figure}[h]
    \centering
    \includegraphics[scale=0.4]{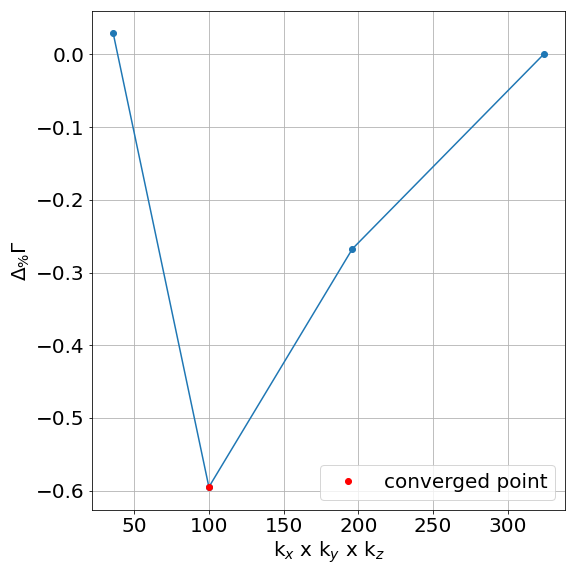}
    \caption{\textbf{k}-mesh convergence for monolayer MoS$_2$.}
\end{figure}
\begin{figure}[h]
    \centering
    \includegraphics[scale=0.4]{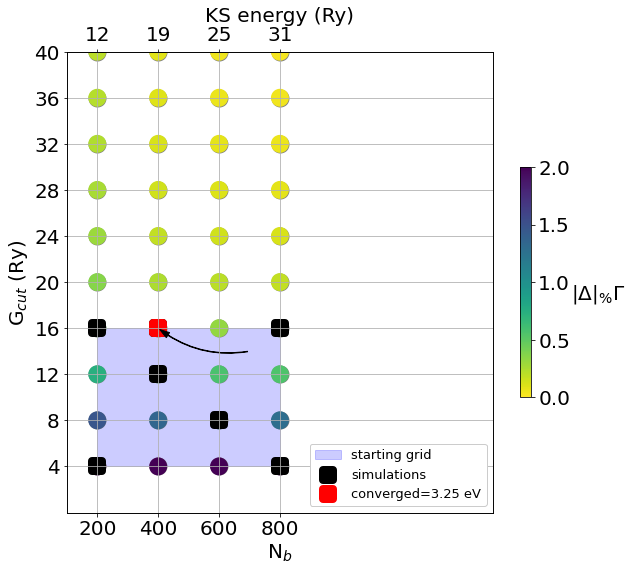}
    \caption{Convergence of $N_b$ and $G_{cut}$ for silicon.}
\end{figure}
\begin{figure}[h]
    \centering
    \includegraphics[scale=0.4]{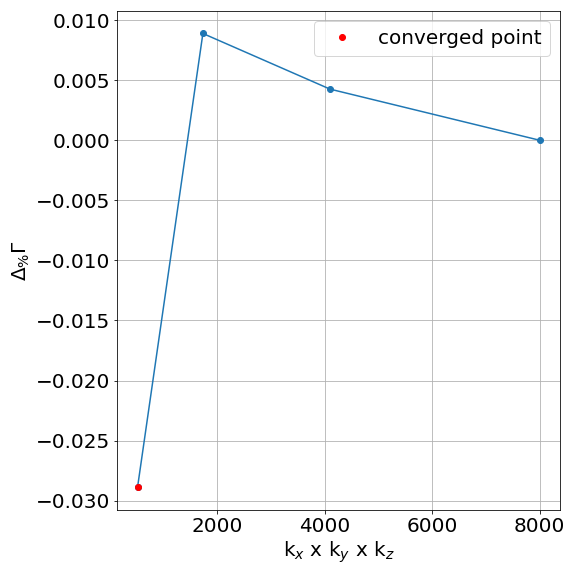}
    \caption{\textbf{k}-mesh convergence for silicon.}
\end{figure}
\begin{figure}[h]
    \centering
    \includegraphics[scale=0.4]{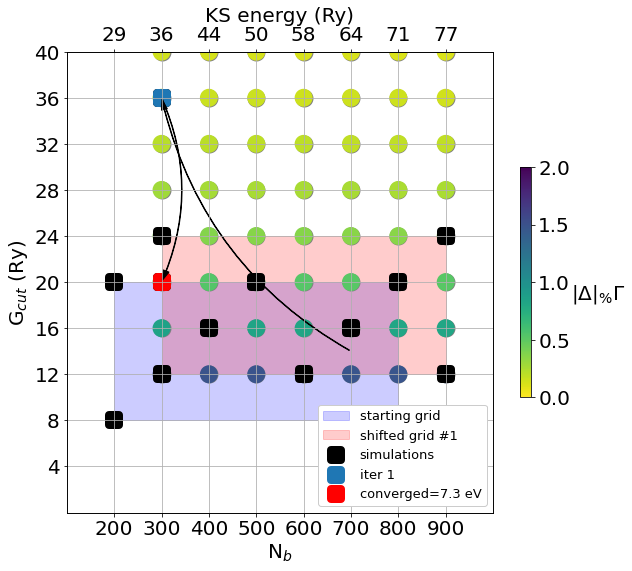}
    \caption{Convergence of $N_b$ and $G_{cut}$ for diamond.}
\end{figure}
\begin{figure}[h]
    \centering
    \includegraphics[scale=0.4]{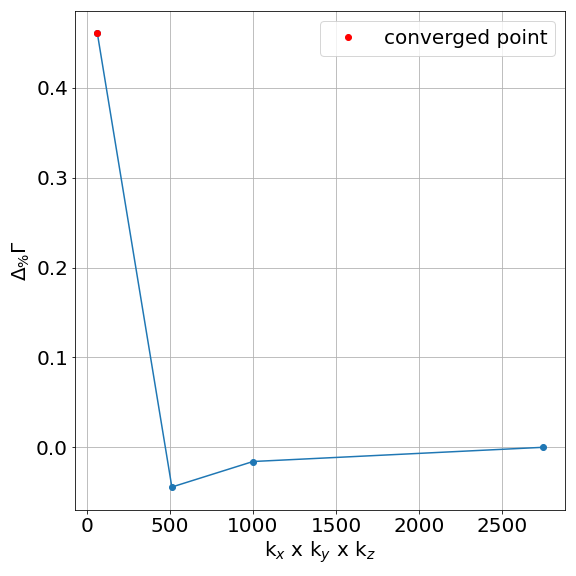}
    \caption{\textbf{k}-mesh convergence for diamond.}
\end{figure}
\begin{figure}[h]
    \centering
    \includegraphics[scale=0.4]{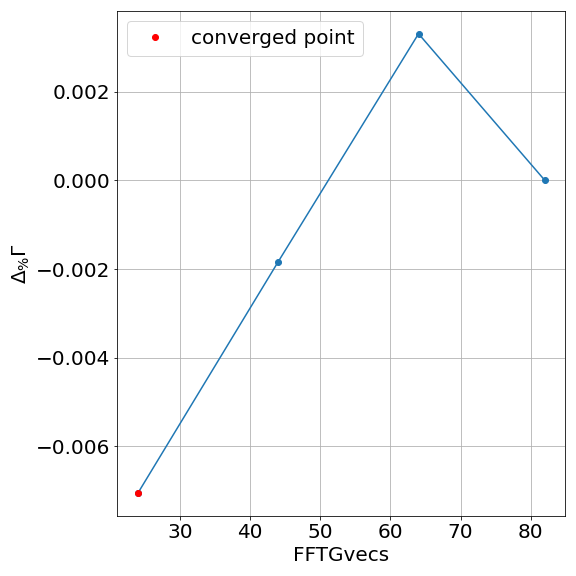}
    \caption{FFT convergence for diamond.}
\end{figure}
\begin{figure}[h]
    \centering
    \includegraphics[scale=0.4]{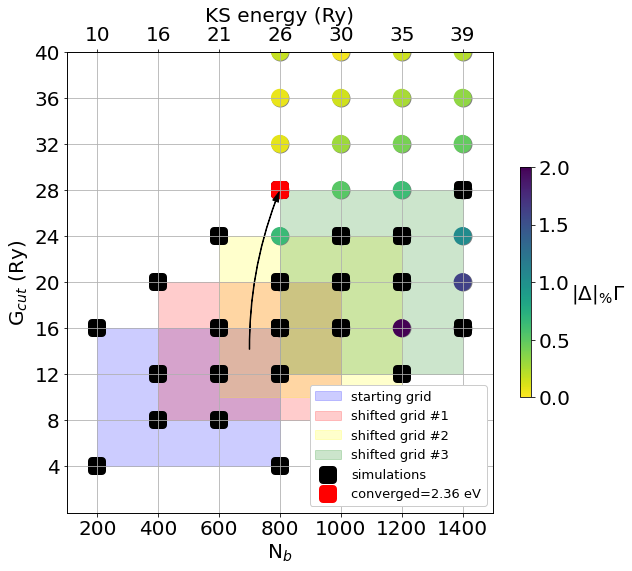}
    \caption{Convergence of $N_b$ and $G_{cut}$ for ZnO.}
\end{figure}

\newpage

\section{Wannierization at the GW level and restart from DFT Wannierization}
\begin{figure*}
  \includegraphics[scale=0.7]{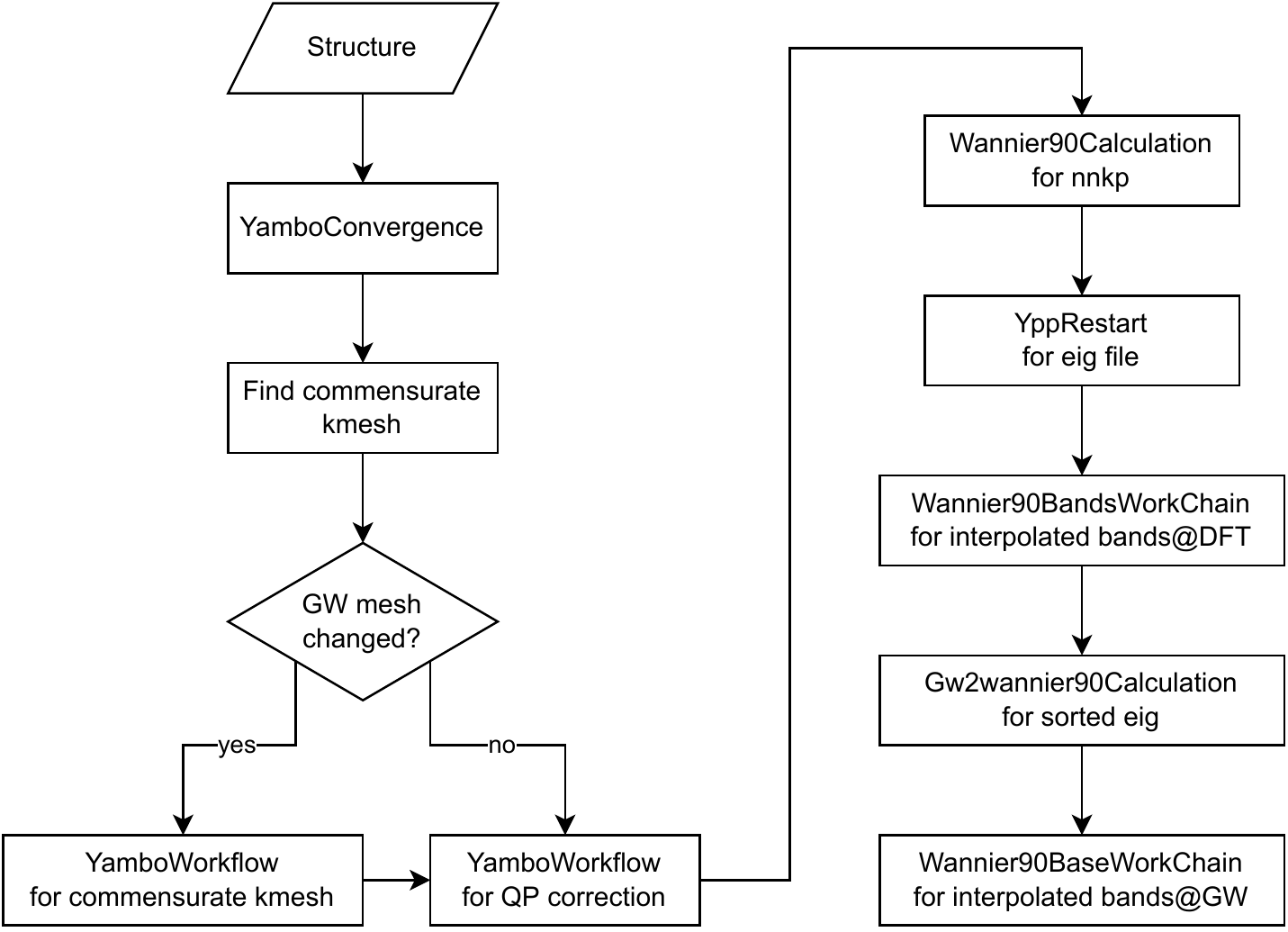}
  \cprotect
  \caption{Detailed flowchart of the \texttt{YamboWannier90WorkChain} for
    automated GW convergence and Wannier interpolated GW band structure.
    The workflow performs the \texttt{Yambo} convergence, searching
    of commensurate kpoint mesh between \texttt{Yambo} and \texttt{Wannier90}, and running the Yambo
    quasiparticle calculation.
    The QP correction is provided and the final steps of the flow comprise the Wannierizations and the band
    interpolations at DFT level and GW level.} 
    \label{fig:gww90_flowchart_SI}
\end{figure*}

Figure
\ref{fig:gww90_flowchart_SI} shows a 
detailed flowchart of the Wannierization workflow
for GW band structure.
The right column shows further details of 
Wannierization with respect to the ones provided in the main text. Firstly, a \texttt{Wannier90Calculation} is launched
to generate the \texttt{nnkp} file for the subsequent \texttt{YppRestart},
which performs the \texttt{ypp} calculation for writing the QP correction in
the \texttt{eig} file format for \texttt{Wannier90}. Then, a
\texttt{Wannier90BandsWorkChain} is launched for Wannier interpolated bands at
DFT level. Next, a \texttt{Gw2wannier90Calculation} sorts the new eigenenergies
in ascending order, and rewrites relevant files. Finally, a
\texttt{Wannier90BaseWorkChain} is launched to obtain the Wannier interpolated
bands at G$_0$W$_0$ level.

Some special care need to be taken at the step of incorporating the GW correction to
the \texttt{Wannier90} inputs, i.e., the initial projection matrices
$A_{mn\bm{k}}$, the overlap matrices $M_{mn\bm{k}\bm{b}}$, the eigenvalues
$\epsilon_{n\bm{k}}$, and the Bloch wavefunctions $\left| u_{n\bm{k}}
  \right\rangle$, where the $\bm{b}$ is the $\bm{b}$ vectors connecting
neighboring kpoints\cite{Pizzi2020}. Since the GW correction might change the
order of eigenvalues, and the Wannierization disentanglement process requires
the eigenvalues to be sorted in ascending order, we need to sort the GW
corrected eigenenergies and rewrite the $A_{mn\bm{k}}$, $M_{mn\bm{k}\bm{b}}$,
and $\left| u_{n\bm{k}} \right\rangle$ files accordingly. Then the
Wannierization and interpolation proceed as usual. Another option is reusing
the unitary transformation matrices $U_{mn\bm{k}}$ from the Wannierization at
DFT level. Since the G0W0 only corrects energies, while the wavefunctions are
left unchanged, it is expected that the orbital characters of the wavefunctions
are unchanged, thus the disentanglement process should still choose the same
set of orbitals, i.e., the previous $U_{mn\bm{k}}$ should work as well for the
G$_0$W$_0$ corrections. Therefore, we can also reuse the $U_{mn\bm{k}}$ for the
Wannier interpolation of G0W0 eigenenergies
thus skipping the second disentanglement step. 
The Wannier interpolation is the
inverse Fourier transform of the Hamiltonian on the kpoint path for the band
structure, the order of eigenenergies is irrelevant, thus the $A_{mn\bm{k}}$,
$M_{mn\bm{k}\bm{b}}$, and $\left| u_{n\bm{k}} \right\rangle$ can be left
intact. This can speed up large-scale calculations where the file size
of these matrices are large, since usually the disk read/write speed is the
slowest during the calculations. Moreover, reusing the $U_{mn\bm{k}}$ means it
will reach the same minimum as the Wannierization at DFT level, thus the
Wannier interpolation accuracy of GW eigenenergies is expected to be similar to that at
DFT level.
%
%
\subsection{Effective band interpolation around given energy windows to reduce computational costs} 
\label{effective_interp_gww90}
The calculation of a large number of quasiparticle corrections can become easily computationally unfeasible for large systems. For example, in the case of PdCoO$_2$~\cite{PdCoO2_strain}, a bulk metallic system of interest for Fermi surface measurements, the minimum number of GW eigenvalues needed to obtain the interpolated band structure is N$_{QP} \sim 3 \cdot 10^4$. We propose here a method to effectively reduce the number of GW evaluation needed, if only a given energy region of interest requires an accurate band interpolation. Within this approximation, the quasiparticle correction to KS eigenvalues is introduced in such a way to be the exact one in a given energy range, and replaced by a scissor and stretching evaluation outside. The exact GW correction goes continuously into the scissored and stretched one, and viceversa, through a Fermi-Dirac smearing, considering an energy window centered at $E_{center}$ and of width $\mu$:
\begin{eqnarray}
\begin{aligned}
    E^{approx.}_{GW} = & \frac{E^{true}_{GW}}{1+e^{\frac{|E_{KS}-E_{center}|-\mu} {T}}} \\
    & + (1-\frac{1}{1+e^{\frac{|E_{KS}-E_{center}|-\mu} {T}}})\times (A*E_{KS}+b)
\end{aligned}
\label{G+S}
\end{eqnarray}
The smearing parameter T is tuned in such a way to reduce the bump between the two different types of corrections.
The $E_{center}$, in terms of KS energies, is user defined. 
\begin{figure}[]
    \centering\includegraphics[width=1\columnwidth]{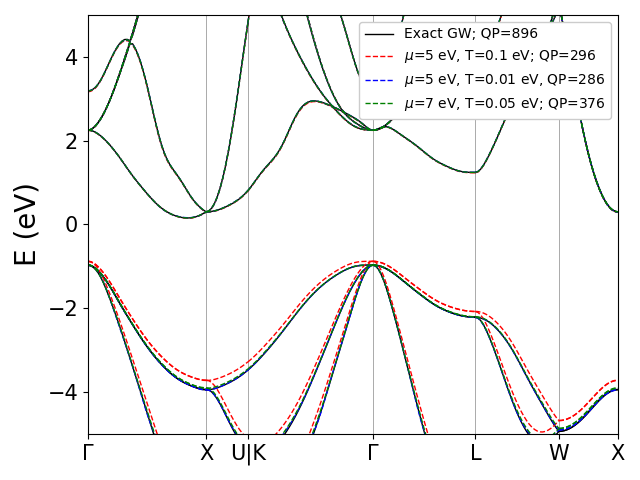}
    \caption{\textbf{FD and scissor corrections for Silicon.} We observe a rather good agreement, around $10^{-2}$ eV, between the exact solution and the approximated ones for T$\leq10^{-2}$ eV. $E_{center}$ here is chosen to be the KS-DFT Fermi level.}
    \label{fig:Silicon_FD}
\end{figure}
In Fig.~\ref{fig:Silicon_FD} are shown the results of the interpolation for different values of $\mu$ and $T$, considering $E_{center}$=$E_{Fermi}^{KS}$. We observe that results are in very good agreement ($\leq 10^{-2}$ eV) for $T\leq10^{-2}$ eV.
In particular for the metals case, we are often interested only in the region around the Fermi level, for example to compute the Fermi surface. This means that we require the exact GW correction to DFT only for a small region belonging to an energy window supposed to include the Fermi level, approximating the corrections for the other eigenvalues in such a way to still have a good Wannier interpolation but reducing the number of explicit GW eigenvalues. Anyway, we stress that since G$_0$W$_0$ is not particle conserving, a large region centered at the DFT Fermi energy is needed in order to possibly include the new and unknown GW Fermi level, that can be calculated a posteriori.
\begin{figure*}[h]
    \centering
    \includegraphics[scale=0.1]{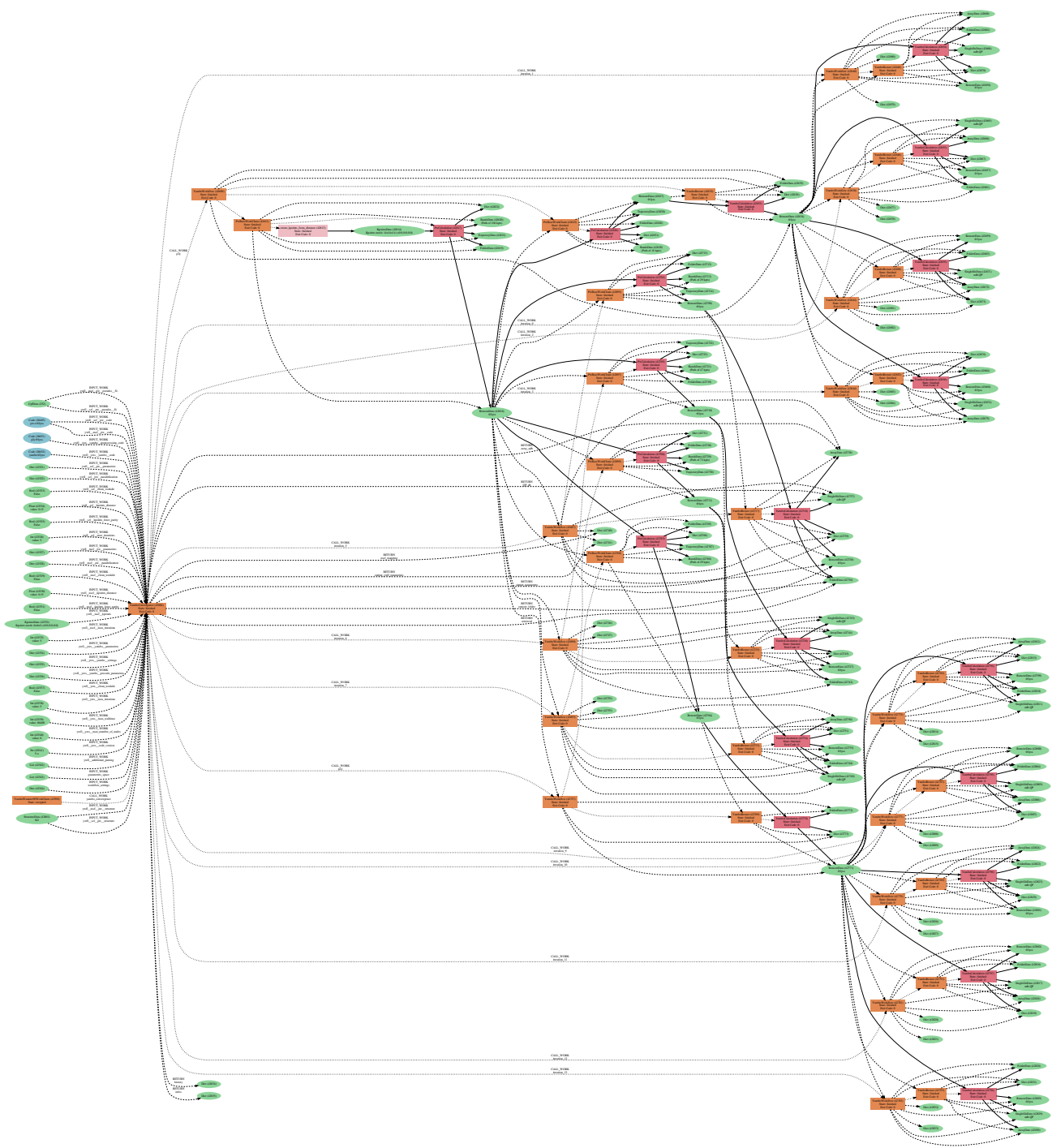}
    \caption{\textbf{Example of provenance graph for \texttt{YamboConvergence}}. Provided the inputs (the first column of boxes from the left), the \texttt{YamboConvergence} calls a series of \texttt{YamboWorkflows} to perform the needed runs of \texttt{PwCalculation} and \texttt{YamboCalculation}, represented by the red boxes. Outputs are then generated and analysed by the workflow.}
\end{figure*}
\bibliography{./bib/bibliography,./bib/codes, ./bib/mbpt_theory}